\documentclass[fleqn,usenatbib]{mnras}

\usepackage{newtxtext,newtxmath}

\usepackage[T1]{fontenc}

\DeclareRobustCommand{\VAN}[3]{#2}
\let\VANthebibliography\thebibliography
\def\thebibliography{\DeclareRobustCommand{\VAN}[3]{##3}\VANthebibliography}

\newcommand{\msun}{\mathrm{M}\textsubscript{\(\odot\)}}
\newcommand{\rsun}{\mathrm{R}\textsubscript{\(\odot\)}}
\newcommand{\Lsun}{\mathrm{L}\textsubscript{\(\odot\)}}
\usepackage{pdflscape}

\usepackage{graphicx}	
\usepackage{amsmath}	
\usepackage{rotating}
\usepackage[dvipsnames]{xcolor}
\usepackage{ulem}

\title[RGB Mass Loss in M4 with Asteroseismology]{Integrated Mass Loss of Evolved Stars in M4 using Asteroseismology}

\author[M. Howell et al.]{
Madeline Howell,$^{1,2}$\thanks{E-mail: madeline.howell1@monash.edu}
Simon W. Campbell,$^{1,2}$
Dennis Stello$^{3,4,5,2}$
and Gayandhi M. De Silva$^{6,2}$
\\
$^{1}$School of Physics and Astronomy, Monash University, Clayton, VIC 3800, Australia.\\
$^{2}$ARC Centre of Excellence for All Sky Astrophysics in 3 Dimensions (ASTRO 3D)\\
$^{3}$School of Physics, University of New South Wales, NSW 2052, Australia.\\
$^{4}$Sydney Institute for Astronomy (SIfA), School of Physics, University of Sydney, NSW 2006, Australia.\\
$^{5}$Stellar Astrophysics Centre, Department of Physics and Astronomy, Aarhus University, DK-8000 Aarhus C, Denmark.\\
$^{6}$Australian Astronomical Optics, Faculty of Science and Engineering, Macquarie University, Macquarie Park, NSW 2113, Australia.\\
}

\date{Accepted XXX. Received YYY; in original form ZZZ}

\pubyear{2022}

\begin{document}
\label{firstpage}
\pagerange{\pageref{firstpage}--\pageref{lastpage}}
\maketitle

\begin{abstract}
Mass loss remains a major uncertainty in stellar modelling. In low-mass stars, mass loss is most significant on the red giant branch (RGB), and will impact the star's evolutionary path and final stellar remnant. Directly measuring the mass difference of stars in various phases of evolution represents one of the best ways to quantify integrated mass loss. Globular clusters (GCs) are ideal objects for this. M4 is currently the only GC for which asteroseismic data exists for stars in multiple phases of evolution. Using \textit{K2} photometry, we report asteroseismic masses for 75 red giants in M4, the largest seismic sample in a GC to date. We find an integrated RGB mass loss of $\Delta\overline{M} = 0.17 \pm 0.01 ~\msun$, equivalent to a Reimers' mass-loss coefficient of $\eta_R = 0.39$. Our results for initial mass, horizontal branch mass, $\eta_R$, and integrated RGB mass loss show remarkable agreement with previous studies, but with higher precision using asteroseismology. We also report the first detections of solar-like oscillations in early asymptotic giant branch (EAGB) stars in GCs. We find an average mass of $\overline{M}_{\text{EAGB}}=0.54 \pm 0.01 ~\msun$, significantly lower than predicted by models. This suggests  larger-than-expected mass loss on the horizontal branch. Alternatively, it could indicate unknown systematics in seismic scaling relations for the EAGB. We discover a tentative mass bi-modality in the RGB sample, possibly due to the multiple populations. In our red horizontal branch sample, we find a mass distribution consistent with a single value. We emphasise the importance of seismic studies of GCs since they could potentially resolve major uncertainties in stellar theory. 
\end{abstract}

\begin{keywords}
Stars: low-mass -- Stars: mass-loss -- Asteroseismology -- Galaxy: globular clusters: individual: NGC 6121 (M4)
\end{keywords}



\section{Introduction}

Mass loss is a defining factor in a star's evolution, having a direct impact on the evolutionary path and the final stellar remnant. However, mass loss rates remain a major uncertainty in stellar modelling. When modelling low-mass stars, the amount of mass lost -- which is most significant on the red giant branch (RGB) -- will influence the evolution of the subsequent horizontal branch (HB) and asymptotic giant branch (AGB) phases. RGB mass loss rates in models are parameterised by stellar properties (such as luminosity ($\mathrm{L}$), radius ($\mathrm{R}$), mass ($\mathrm{M}$), surface gravity ($\log\mathrm{g}$), and effective temperature ($T_{\text{eff}}$)), and are implemented by simple relations such as the Reimers' scheme \citep{Reimers_massloss_rate}, or the recent adaptation, the Schr\"oder \& Cuntz scheme \citep[][]{Schroder_cunts_massloss_rate}. Both of these empirical mass-loss rates are dependent on free scaling parameters, $\eta$, which need to be calibrated, introducing uncertainty into the rates.

It has been proposed that significant mass loss in low-mass stars does not occur until the RGB bump \citep{bumpmasslosstheory1,massloss_bump_3,massloss_bump_4}. This theory deviates from the Reimers' and Schr\"oder \& Cuntz schemes, which assume mass loss over the entirety of the RGB phase, with the largest mass loss rate towards the RGB tip. There are also suggestions in the literature that the RGB mass loss rate is close to constant across this phase \citep{massloss_obs2}, or in contrast, episodic \citep{episodic_massloss}. The disagreements between these theories are due to a lack of physical understanding of the RGB mass loss process (e.g. \citealt{mullan2019}). 

One way to quantify the amount of mass loss in low-mass stars is to use globular clusters (GCs). 
GCs are ideal objects for this because they contain stars with similar metallicities and ages, and are found at various evolutionary phases. With a typical GC turn-off mass of $\sim0.8~\msun$, their HB stars have masses\footnote{We note that the position in the colour-magnitude diagram (CMD) of HB stars is sensitive to mass (see eg. the review by \citealt{Catelan_HB_review}).} of $\sim0.6~\msun$ (mass estimates for both of these phases are generally determined from CMD fitting with isochrones), and their white dwarves have masses of $\sim0.50-0.55~\msun$ \citep[][determined from Balmer line fitting]{GC_WD_masses}. This results in an expected integrated RGB mass loss of $\Delta M \sim0.2~\msun$, which is reproduced by stellar models calibrated to photometry of GCs \citep[e.g.][]{Macdonald_omega_centauri_massloss,tucan_47}. 

An important point when determining integrated RGB mass-loss from RGB/HB mass differences, is that there is an \textit{initial} mass difference between the RGB and HB stars we observe today. The current HB stars had higher initial masses than the current RGB stars, and hence have evolved faster (indeed this is why we observe stars in different phases of evolution in GCs). This mass difference needs to be factored in when calculating integrated mass loss (see eg. Figs 6 and 7 in \citealt{2012NGC67916819}, for an open cluster case). In GCs this initial mass difference is negligible ($\sim 10^{-3}~\msun$; PARSEC isochrones; \citealt{bressan2012}) since it is below the precision of our asteroseismic mass measurements ($\sim 10^{-2}~\msun$), and well below the precision of other methods. Thus for the current study, we consider the initial masses of the RGB and red HB (RHB) to be equal.

Mass loss of red giants in GCs has been previously studied using a variety of techniques. For example, \citet{dusty_massloss1, infrared} used infrared photometry of Tuc 47 to detect dust in the circumstellar envelopes of stars along the entire RGB phase, to estimate an empirical mass loss rate. However, the existence of significant dust production in low luminosity RGB stars ($L<1000~\Lsun$) has been questioned by \citet{nodustproduction1} and \citet{nodustproduction}. They report that dust production for these stars will only occur $\sim1$~mag from the RGB tip, and that the mass loss rates derived in \citet{dusty_massloss1} over-estimate the mass loss in the RGB phase. Other observations of mass loss use H$\alpha$ emission to trace mass motions near the stellar surface \citep[e.g.][]{halpha1,halpha2,halpha3,halpha4,massloss_obs2}. However, it has been suggested that the H$\alpha$ wings could naturally arise in the stellar chromosphere \citep{Dupree1,Dupree2}, and may not be an indicator of mass loss. 

Given the caveats and uncertainties of the aforementioned techniques, measuring the mass difference between evolutionary stages would be a much more direct and reliable approach to quantify the integrated mass loss.

When determining the total mass lost in the evolution of stars in GCs, we must also consider the existence of multiple populations \citep[as observed by numerous studies; e.g.][]{Mulpop_obs4,Mulpop_obs1, Mulpop_obs3}. These multiple populations are defined by their light elemental abundances \citep{light_elemental_abundances1,light_elemental_abundances2}, where the inferred helium variation can result in different evolutionary paths, and therefore different current observed stellar masses between the populations \citep{chloes_paper, 2pops_models}. Furthermore, it has been shown that the mass loss rates between the sub-populations may be different \citep{massloss_2pops}. If true, this implies that there should be an enhanced mass difference on the HB, because this phase shows the cumulative effect of the RGB mass loss histories. 

With the ability to accurately estimate mass using stochastically-excited pressure-mode oscillations, also known as solar-like oscillations, asteroseismology can be used very successfully to study stellar evolution. Solar-like oscillations in a star's power spectrum can be characterised by two global seismic quantities, $\nu_{\text{max}}$, the frequency of the maximum acoustic power, and $\Delta\nu$, the large frequency spacing between adjacent overtone oscillation modes. Both of these seismic quantities are correlated with fundamental stellar properties; $\nu_{\text{max}}$ is related to the surface gravity and temperature, $\nu_{\text{max}} \propto gT_{\text{eff}}^{-1/2}$, and $\Delta\nu$ is related to the mean density of the star, $\Delta\nu \propto \overline{\rho}^{1/2}$ \citep{scaling_relation1, scaling_relation2, scaling_relation3}. From these relations and the Stefan-Boltzmann luminosity law, $L\propto R^2T_{\text{eff}}^4$, four seismic mass equations can be derived :
\begin{align}
\label{eq:mass_relation}
    &\left(\frac{M}{M_{\odot}}\right)\simeq\left(\frac{\nu_{\text{max}}}{\nu_{\text{max},\odot}}\right)^3\left(\frac{\Delta\nu}{\Delta\nu_{\odot}}\right)^{-4}\left(\frac{T_{\text{eff}}}{T_{\text{eff},\odot}}\right)^{3/2}\\
\label{eq:mass_relation2}
    &\left(\frac{M}{M_{\odot}}\right)\simeq\left(\frac{\Delta\nu}{\Delta\nu_{\odot}}\right)^{2}\left(\frac{L}{L_{\odot}}\right)^{3/2}\left(\frac{T_{\text{eff}}}{T_{\text{eff},\odot}}\right)^{-6}\\
\label{eq:mass_relation3}
    &\left(\frac{M}{M_{\odot}}\right)\simeq\left(\frac{\nu_{\text{max}}}{\nu_{\text{max},\odot}}\right)\left(\frac{L}{L_{\odot}}\right)\left(\frac{T_{\text{eff}}}{T_{\text{eff},\odot}}\right)^{-7/2}\\
\label{eq:mass_relation4}
    &\left(\frac{M}{M_{\odot}}\right)\simeq\left(\frac{\nu_{\text{max}}}{\nu_{\text{max},\odot}}\right)^{12/5}\left(\frac{\Delta\nu}{\Delta\nu_{\odot}}\right)^{-14/5}\left(\frac{L}{L_{\odot}}\right)^{3/10}
\end{align}
This provides a direct way of calculating masses for individual stars and testing for systematics in each of the four mass scales.

Deriving integrated mass loss using asteroseismology was first attempted on the open clusters NGC6791 and NGC6819  \citep{2012NGC67916819, 2017NGC6819} using \textit{Kepler} photometry \citep{kepler1}. An integrated mass loss on the RGB was estimated by calculating the difference between the average RGB and red clump (core helium-burning star) masses, and comparing to isochrones. NGC6791 and NGC6819 were found to have small mass differences of $\Delta\overline{M}_{6791} = 0.09\pm 0.03$(random)$\pm0.04$(systematic)$~\msun$ and $\Delta\overline{M}_{6819} = -0.03\pm0.04~\msun$. \citet{M67_stello} also reported a small RGB mass difference of $0.02-0.05~\msun$ for the open cluster M67 using photometry from \textit{Kepler's} second mission, \textit{K2} \citep{K2_mission}. The isochrones in the OC studies showed that the initial mass differences between the HB and RGB stars were of order $10^{-2}~\msun$, a similar size to the uncertainties, and in some cases to the mass difference itself. Hence, the small resulting integrated mass losses found by these studies were attributed to the young ages of these clusters, which coincides with low predicted mass loss according to the Reimers' formulation.
%
%
%
In contrast, GC stars are expected to present large mass differences, because of their long lifetimes and relatively homogeneous initial masses.

The field of view along the ecliptic of the \textit{K2} mission provided the chance to observe some GCs. One cluster, M4 (NGC 6121), provided the only opportunity to seismically determine an integrated mass loss on the RGB, since the HB is bright enough to detect solar-like oscillations. M4 is known for being one of the closest GCs, making it relatively bright, and it has an age of ~$11-12$ Gyrs \citep{M4_age1,2nd_M4_mass_study} and a metallicity of [Fe/H]~$\approx -1.1\pm0.07$ \citep{m4_metallicity,M4_metallicity2, chloes_paper}. Although the \textit{K2} pixel scale of 4''/pixel is not entirely optimised for the dense fields of GCs, it provides an excellent science opportunity to study the mass loss of evolved stars using asteroseismology. 

An initial attempt of performing asteroseismology on M4 stars was conducted from ground-based photometry \citep{M4_ground}, but no signals were detected. Similarly, \citet{seismo_NGC6397}, who used \textit{Hubble Space Telescope} photometry of the GC NGC6397 to find evidence of solar-like oscillations, were not able to derive useful measurements of $\nu_{\text{max}}$ and $\Delta\nu$. Since the release of the \textit{K2} data, there has been one seismic study on M4 \citep[][hereby M16]{Miglio_M4_study}. Solar-like oscillations were detected in 8 K-giants (7 RGB stars and 1 RHB star) and stellar mass estimates were calculated using the seismic relations (Eq.~\ref{eq:mass_relation}-\ref{eq:mass_relation4}). A final mean mass for the RGB sample of $0.84~\msun$ was reported by M16, and an average mass and scatter of $0.68\pm0.12~\msun$ was found for the RHB star. The integrated RGB mass loss cannot be reliably estimated in the M16 study, since only one RHB star (with a large mass uncertainty) was sampled.

In the current study, we substantially increase the number of M4 evolved stars with detected solar-like oscillations. The increase in sample size allows us to reduce the uncertainties on the mean masses in each phase of evolution, thereby measuring a precise value for the integrated mass loss. To achieve this we completed our own membership study and used our custom detrending pipeline for \textit{K2} data to estimate masses for stars in the RGB, RHB and early AGB (EAGB; the phase of evolution directly after the HB) evolutionary phases. In Section~\ref{sec:data_prep} we describe our methodology of target selections and the detrending of \textit{K2} photometric data. In Sections~\ref{sec:seismo} and \ref{sec:stellar_params} we outline how we measured the global asteroseismic parameters and the determination of other stellar characteristics necessary to estimate seismic masses. In Sections~\ref{sec:mass} and \ref{sec:discussion}, we present and discuss our mass results and their implications.

\label{sec:intro}

\section{Data Preparation}
\label{sec:data_prep}
\subsection{Target sample selection}

\begin{figure*}
	\centering
	\includegraphics[width=1.5\columnwidth]{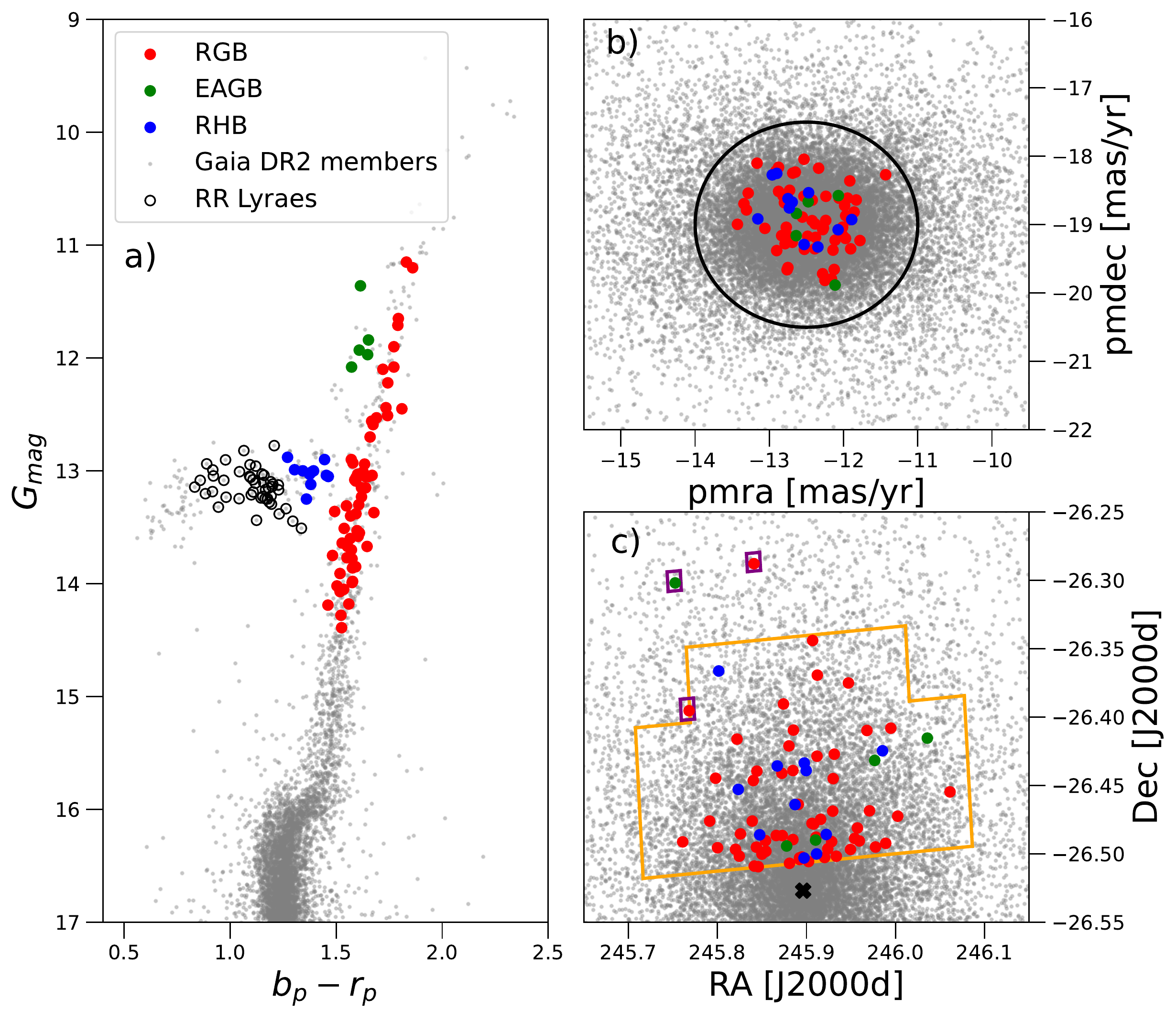}
	\caption{\textbf{a) } CMD of the final M4 sample using Gaia DR2 photometry (grey). Stars for which we were able to detect solar-like oscillations are classified into three evolutionary phases; RGB, RHB and EAGB, and are indicated by the larger coloured points. To demonstrate the positions of the instability strip in the M4 CMD, RR Lyrae stars identified by \citet{RRLyrae_cat} are indicated by black open circles. No reddening corrections have been made on this photometry. \textbf{b) } Gaia proper motions of the same sample. This diagram was the initial step in member classifications. Stars within a circle of $1.5~\mathrm{mas/yr}$ centred at $(-12.5, -19.0)~\mathrm{mas/yr}$, illustrated as a black circle, were included as potential members. \textbf{c) } The spatial positions of the sample relative to the cluster centre (black cross). We also show the M4 \textit{K2} superstamp (orange) and individual target pixel files (purple) for the stars EPIC203390179/M4AGB62, EPIC203362169/M4RGB225, and EPIC203394211/M4RGB408.}
	\label{fig:CMD}
\end{figure*}

We established an initial M4 sample from the Gaia DR2 catalog \citep{GAIA_Dr2_1, GAIA_dr2_2} with Gaia magnitude $G_{\text{mag}} < 15$ mag (to ensure targets are sufficiently bright to have detectable oscillations) and within the field of view of the M4 \textit{K2} superstamp (comprised of 16 target pixel files). Targets were classified into various evolutionary phases: RGB, RHB, blue HB, EAGB, and blue stragglers, based on their positions in the Gaia DR2 photometric CMD (Fig. \ref{fig:CMD}a). We then assigned our own star identifications using the naming convention as follows: GC name (M4) + evolutionary phase (eg RGB, RHB or AGB) + star number (starting at 1). 

Cluster membership was derived from Gaia DR2 proper motions (Fig. \ref{fig:CMD}b), where stars were included in our sample if they were within a circle of radius $1.5~\mathrm{mas/yr}$ and centred at $(-12.5, -19)~\mathrm{mas/yr}$, in the proper motion right ascension and declination direction respectively. Our membership sample was cross checked with the \citet{GAIA_EDR_GCs} GC catalog, which calculated membership probabilities based on Gaia EDR3 parallaxes and proper motions. We found that all our stars, except M4AGB01 and M4RGB169, had a membership probability metric of $\geq 0.99$. The star M4AGB01 was contained in the \citet{GAIA_EDR_GCs} catalog, but they rejected this star because it had a much larger parallax than the M4 average. Upon inspection, M4AGB01 had a large Gaia EDR3 \texttt{RUWE} goodness-of-fit statistic of $4.03$, which could be indicative of unreliable astrometry or binarity. However, because the star M4AGB01 has been classified as a member using radial velocities and chemical abundances in \citet[][hereby MacLean18]{chloes_paper}, we included it in our study. The star M4RGB169 was not contained within the \citet{GAIA_EDR_GCs} catalog. However, it was also included in the MacLean18 spectroscopic study and classified as a member. We also found three member stars not contained in the M4 superstamp, which had their own individual \textit{K2} target pixel files (EPIC203390179/M4AGB62, EPIC203362169/M4RGB225, \& EPIC203394211/M4RGB408). These stars have also been included in our sample. 

\subsection{Photometric detrending pipeline}
Only about half of M4 was observed in Campaign 2 of the \textit{K2} mission (Fig. \ref{fig:CMD}c), with $3856$ 30-minute long cadences over the $\sim80$ day observing period. Some cadences were missing data due to systematic events, such as thruster firing to roll the telescope. Each target pixel file consists of $50 \times 50$ pixels. Due to the large pixel size, there is evidence of blending between stars, especially close to the centre of the cluster. We used a combination of automated aperture masks and custom individual masks to ensure there was no photometric contamination from neighbouring stars. The automated aperture masks were generated by including pixels that were within a threshold level of the brightest pixel over time (assumed to be the centre of the star). After inspection, these aperture masks were then individually adjusted to find the optimal mask for the star, based on the combined differential photometric precision metric \citep[for further details see][]{cdpp_noise}.   

\begin{figure}
	\centering
	\includegraphics[width=\columnwidth]{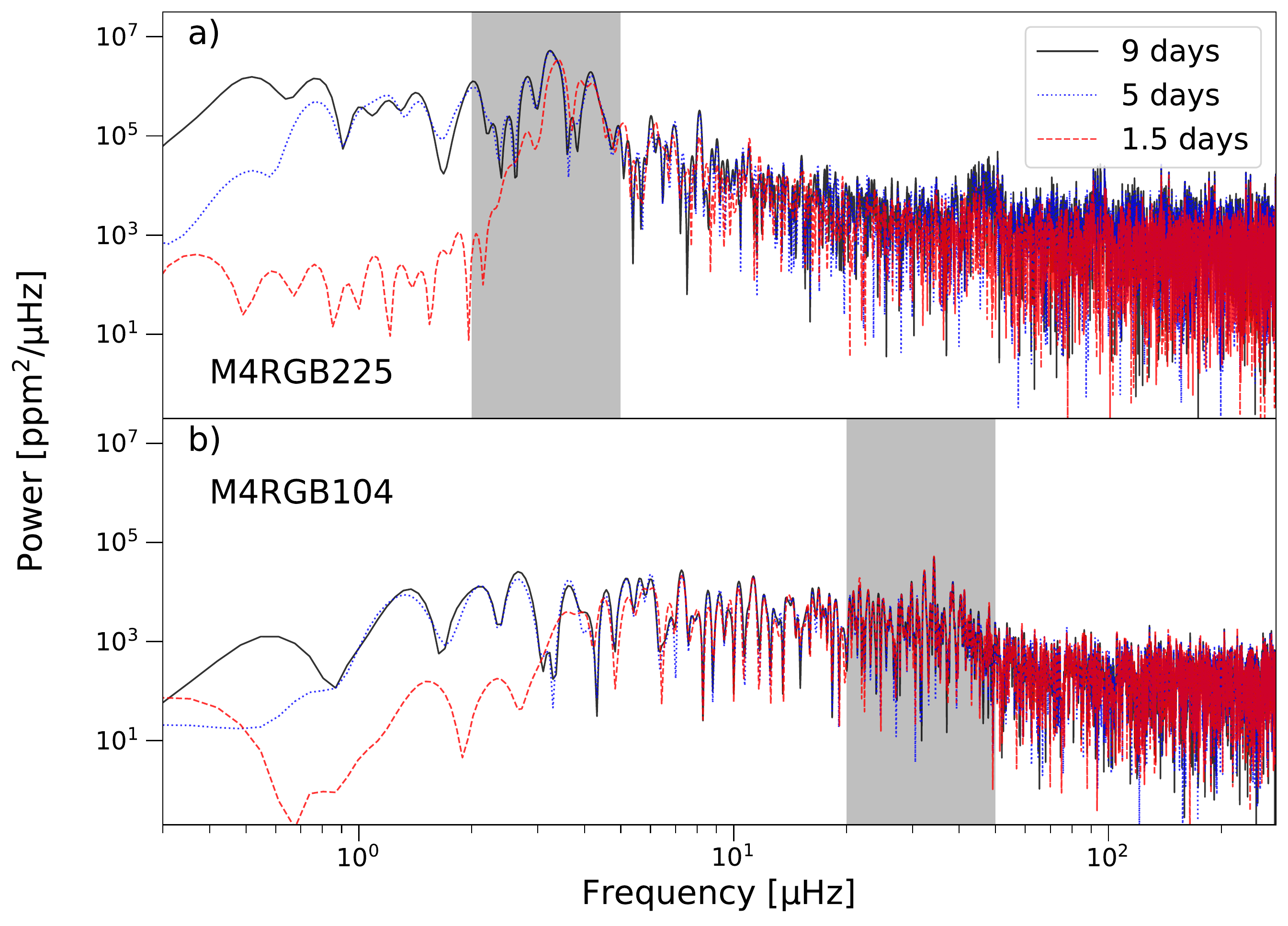}
	\caption{\textbf{a)} Comparison of power spectra for M4RGB225 with three different SFF timescales. The grey shaded area illustrates the envelope of the oscillation power excess for the star. \textbf{b) } Same but for M4RGB104 star with a larger $\nu_{\text{max}}$ similar to the 6-hr roll noise.}
	\label{fig:ts_comp}
\end{figure}

We developed a pipeline using the \texttt{Python} package, \texttt{LightKurve}, \citep{Lightkurve} to detrend the time series in two stages; stage 1 used a pixel-level decorrelator \citep[][]{PLD_creator} on the target pixel file itself and stage 2 used a self-flat fielding corrector (SFF) on the light curve \citep[][]{SFF_cleaning}. The pixel-level decorrelation method was developed for the \texttt{EVEREST} pipeline \citep{PLD_Everest}, and it removes systematic errors due to instrumental drifts from individual pixels. The SFF detrender then removed long-term trends to produce a finalised flattened light curve. For this detrender, we used a timescale of 9 days, which is significantly larger than the typical 1.5 days that is used in the literature \citep[e.g.][]{M67_stello,Wallace_RRLyrae}. In our tests, we found that at smaller timescales the low frequency end of the spectrum was attenuated. This will affect the background fitting to the spectrum, and also the measurement of the global seismic parameters (see Section~\ref{sec:seismo}), for evolved RGB and EAGB stars with low $\nu_{\text{max}}$ ($\nu_{\text{max}}\lesssim 5~\mu\mathrm{Hz}$). By increasing the timescale in the SFF detrender, we retained the power at lower frequencies, while not affecting the power for the rest of the spectrum. To illustrate this, Figure~\ref{fig:ts_comp} shows the power spectra for two stars calculated with different timescales in the SFF detrender; M4RGB225 -- an evolved RGB star with a $\nu_{\text{max}}\approx 3~\mu\mathrm{Hz}$ and M4RGB104 -- a star with a $\nu_{\text{max}}\approx35~\mu\mathrm{Hz}$ similar to the 6-hr roll noise inherent in the \textit{K2} photometry. For the star M4RGB225, a power excess can only be clearly detected when using the longer timescales. Because the SFF detrender was developed to remove systematics such as the 6-hr roll noise, we checked that any noise present was being efficiently removed at the longer timescale. Fig.~\ref{fig:ts_comp}b shows that varying the timescales for the star M4RGB104 does not alter the power excess of the solar-like oscillation. 

Within our pipeline, we restricted the time series to 36 days (cadences 95717-97490) for stars with expected\footnote{The expected $\nu_{\text{max}}$ was determined by making a first pass of our data, and deriving a relation between Gaia DR2 $G_{\text{mag}}$ and $\nu_{\text{max}}$ ($\nu_{\text{max}} \approx 1.78\times10^{-15} G_{\text{mag}}^{14.6} $) for stars showing clear oscillations.} $\nu_{\text{max}} > 20~\mu \mathrm{Hz}$. This step excluded noisy cadences, for example, the `unflagged pointing excursion' at the start of the campaign where there is an extreme flux drop. We did not do this for the stars with predicted $\nu_{\text{max}} \leq 20~\mu \mathrm{Hz}$, because they require higher frequency resolution to properly resolve the oscillation excess power. We instead split the time series of these lower $\nu_{\text{max}}$ stars into two, corresponding to the change in the telescope's roll direction, which occurred about halfway through the campaign. Each section was detrended individually and recombined at the end. Lomb-Scargle power spectral density periodograms \citep{Lomb,Scargle} were then calculated for each star. To determine the efficacy of the data detrending, we used a white noise metric (calculated as the median power between $260-280~\mu\mathrm{Hz}$; \citealt{Stello_2015}) and obtained a range of values between $\sim50-1000~\mathrm{ppm}^2/\mu\mathrm{Hz}$. This is similar to what was achieved in the M16 study, but larger than \textit{K2} field stars of similar magnitudes \citep{Stello_2015}. The larger noise levels of the M4 data could be because of the light contamination from neighbouring stars. This will have a direct impact on the global asteroseismology analysis and the final mass uncertainties. 

Power spectra were manually searched for potential solar-like oscillations. Criteria based on the power spectra were introduced to determine which stars had clear solar-like oscillations or not. A star was removed from the initial membership sample if:
\begin{enumerate}
    \item no power excess was observed in the spectrum. For example, all blue HB and blue straggler stars were removed from the sample, due to the inability to detect solar-like oscillations in their power spectra. This included bright red giants ($G_{\text{mag}} < 10$ mag) with expected $\nu_{\text{max}} \leq 2~\mu\mathrm{Hz}$. This $\nu_{\text{max}}$ is too low to be resolved properly with the short observing period of this \textit{K2} campaign. 
    \item it was a horizontal branch star that was classified as a RR Lyrae by \citet{RRLyrae_cat}, or had an observed RR Lyrae-like spectrum.
    \item it had an ambiguous detection of a solar-like oscillation, typically due to a low signal-to-noise ratio (SNR) $\leq5$. These targets were classified as tentative detections instead (see Table \ref{tab:tentative_detections}), and will not be used in our analysis. Note that the majority of the tentative detections are faint RGB stars. This is expected since the amplitude of the power excess decreases with increasing magnitude on the RGB, resulting in lower SNR.
\end{enumerate}
After applying these criteria we retained a final sample of 75 stars with solar-like oscillations: 59 RGB, 11 RHB, and 5 EAGB. Our final sample is indicated by the coloured points in Figure \ref{fig:CMD}.

\begin{table}
\centering
\footnotesize
\caption{The stars with tentative solar-like oscillation detections in the M4 sample. Because of the lower SNR in their spectra, we were unable to conclusively determine whether they were genuine detections. A predicted $\nu_{\text{max}}$ is given for each star using the estimation method detailed in text. These stars are not used in our analysis.}
    \begin{tabular}{lccc}
      \hline 
      ID & Gaia DR2 ID & $\mathrm{G}_{\text{mag}}$ & $\nu_{\text{max, pred}}$ ($\mu\mathrm{Hz}$)\\
      \hline 
        M4AGB60 & 6045490898082696448 & 11.73 & 7 \\
        M4RGB38 & 6045466743185397888 & 14.5 & 150 \\
        M4RGB119 & 6045466296508858752 & 14.12 & 84 \\
        M4RGB161 & 6045477635223138432 & 12.3 & 15 \\
        M4RGB184 & 6045466738874761472 & 14.57 & 170 \\
        M4RGB187 & 6045490039089161472 & 14.60 & 200 \\
        M4RGB220 & 6045490039089161472 & 14.40 & 160 \\
        M4RGB371 & 6045465643673514624 & 14.72 & 200 \\
        M4RGB397 & 6045490279607362304 & 14.64 & 200 \\
      \hline 
    \end{tabular}
\label{tab:tentative_detections}
\end{table}

\section{Global Asteroseismic Quantities}
\label{sec:seismo}

Using the resultant light curves and power spectra, $\nu_{\text{max}}$ and $\Delta\nu$ were measured using the \texttt{pySYD} pipeline \citep{pySYD_chontos}, which is an adaptation of the \texttt{SYD} pipeline \citep{SYD_huber} written in the \texttt{Python} language. The \texttt{pySYD} pipeline uses optimised Lorentzian-based models for background fitting and heavy smoothing of the power spectrum to estimate $\nu_{\text{max}}$. To measure $\Delta\nu$, \texttt{pySYD} uses an auto-correlation function. The pipeline estimates uncertainties using a Monte Carlo sampling routine. This routine perturbs the power spectrum with stochastic noise. The background of the new perturbed spectrum is then fitted again, and new global seismic parameters are estimated. This is repeated 200 times for each star. 

\begin{figure*}
	\centering
	\includegraphics[width=\textwidth]{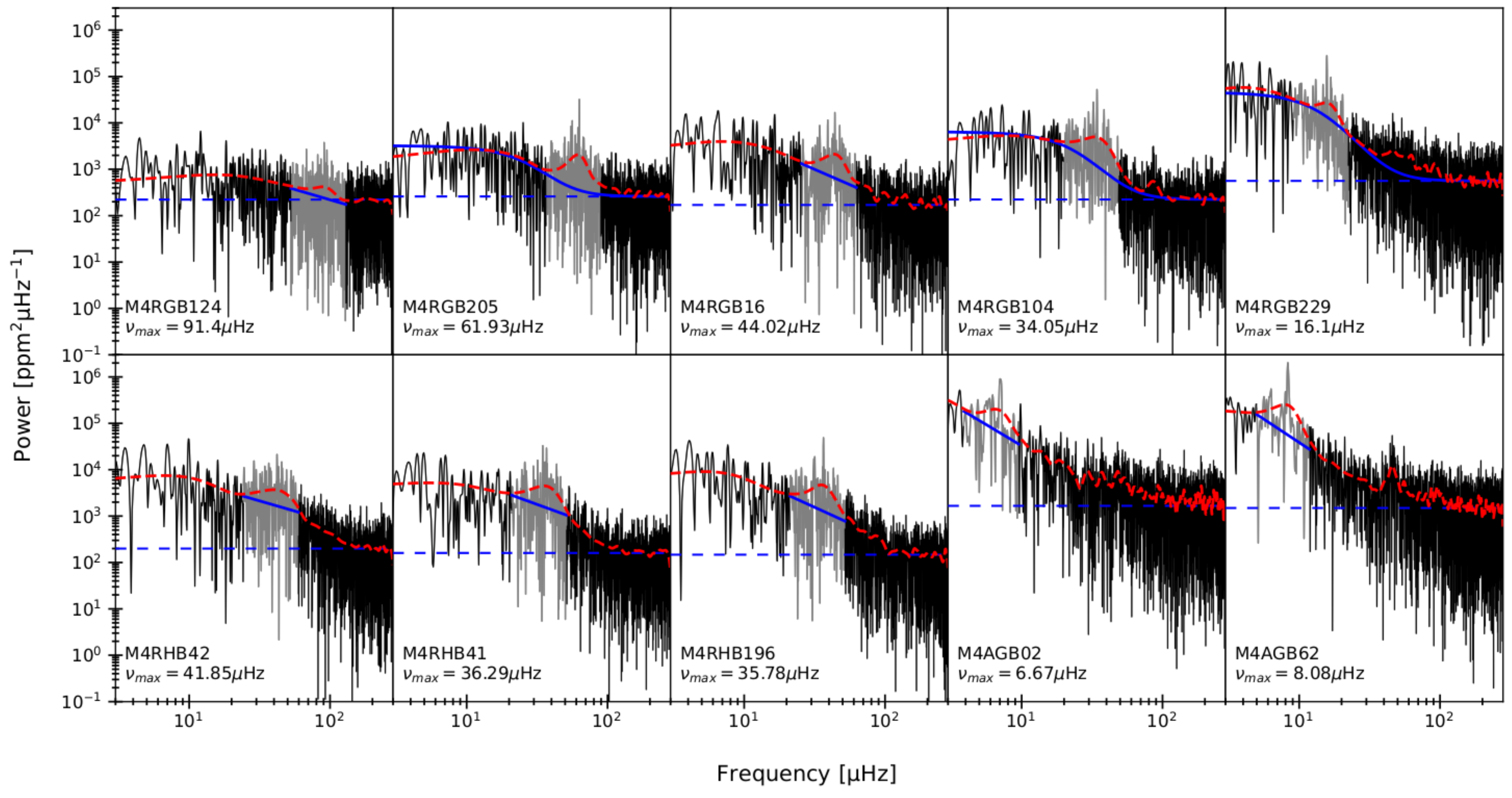}
	\caption{Power spectra using \textit{K2} data for 5 RGB, 3 RHB and 2 EAGB stars (black). Each spectrum is labelled with the star ID and the measured $\nu_{\text{max}}$. The envelope of the power excess for each is represented in grey. Background fits are shown in blue, either the Lorentzian-based or linear fits (solid) and the white noise (dashed). When a linear background model is used, it is only plotted within the power excess envelope. The smoothed power spectrum from which $\nu_{\text{max}}$ was measured is also included in red.}
	\label{fig:PSD_grid}
\end{figure*}

The generally low SNR power excesses meant there were often instabilities in the error estimation routine for the background fitting, causing it to fail for a large proportion of our sample (54 stars). For these stars, we instead used a more stable linear background model for the power excess envelope. The performance of the linear background model was compared to the results of Lorentzian-based models for stars without fitting issues. No systematic difference was found between the resultant global seismic parameters. Examples of background fits for 10 stars in increasing evolution are shown in Fig. \ref{fig:PSD_grid}.

The $\Delta\nu$ estimation method in \texttt{pySYD} requires a prior estimate for this parameter, which is determined by a $\Delta\nu$-$\nu_{\text{max}}$ scaling relation. \citet{deltanu_numax_relation} showed that $\Delta\nu$ is correlated to $\nu_{\text{max}}$ in the form $\Delta\nu = \alpha\nu_{\text{max}}^{\beta}$, where $\alpha$ and $\beta$ are constants calibrated from large samples of stellar seismic data. Due to the mass dependency in the $\Delta\nu$-$\nu_{\text{max}}$ relation, and the lower masses of M4 stars compared to the average field star, the standard $\Delta\nu$-$\nu_{\text{max}}$ relation adopted by \texttt{pySYD} is not optimal for our purpose. We therefore derived our own relation for each evolutionary phase using low-mass and low metallicity stars from four individual seismic studies \citep{Stello_catalog,Vrard_catalog,Yu_study, Dreau_catalog}. Where $\nu_{\text{max}}$ or metallicity was not available in their studies, the APOKASC-2 catalog \citep{APOKASC-2} values were used. These samples only contained RGB and core-helium burning stars, which had been classified accordingly. As the EAGB stars in our M4 sample have similar structures to the RGB stars, we used the RGB sample to calibrate this relation. 

The final empirical $\Delta\nu$-$\nu_{\text{max}}$ scaling relations derived for each evolutionary phase were:
\begin{align}
\label{eq:RGB_sr}
    &\Delta\nu_{\text{RGB}} = 0.3\nu_{\text{max}}^{0.75}\\
\label{eq:RHB_sr}
    &\Delta\nu_{\text{RHB}} = 0.3\nu_{\text{max}}^{0.86}\\   
\label{eq:AGB_sr}
    &\Delta\nu_{\text{EAGB}} = 0.3\nu_{\text{max}}^{0.77}
\end{align}

\begin{figure}
	\centering
	\includegraphics[width=\columnwidth]{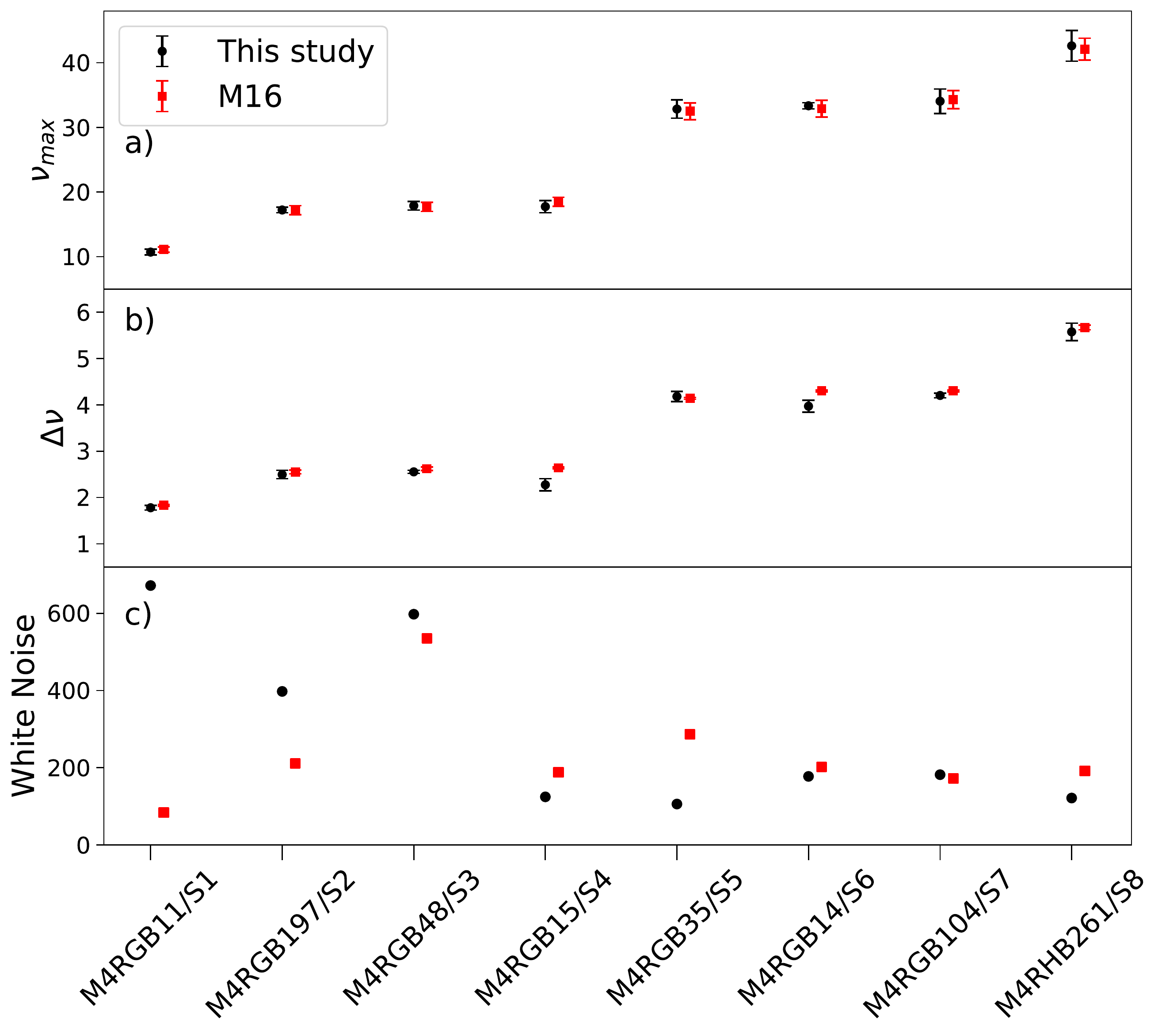}
	\caption{A comparison between seismic results from M16 (red) and our study (black) for the eight  overlapping stars. \textbf{a)} Comparison of $\nu_{\text{max}}$. These estimations are consistent within 1$\sigma$ errors. \textbf{b)} Comparison of $\Delta\nu$ where the results are consistent within 2$\sigma$ errors. \textbf{c)} Comparison of white noise metric. Most stars had the same order of magnitude in noise except for M4RGB11 where we report a much larger noise metric compared to M16.}
	\label{fig:M16_comp}
\end{figure}

As a check, we compared our global seismic parameters and white noise estimates to the 8 stars analysed by M16 (Fig. \ref{fig:M16_comp}). All $\nu_{\text{max}}$ and $\Delta\nu$ values were within the 2$\sigma$ errors, confirming the detections of the solar-like oscillations for these stars. Comparing the magnitude of the uncertainties for $\Delta\nu$, our errors are substantially larger than reported by M16. Most stars had less or similar white noise in their power spectra, except for M4RGB11 (S1 in M16 study), where we found a significantly larger noise metric. However, we note that the larger white noise metric did not impact the measurement of the seismic parameters for this star, which is shown by the consistent values of $\nu_{\text{max}}$ and $\Delta\nu$. 

\begin{figure}
	\centering
	\includegraphics[width=\columnwidth]{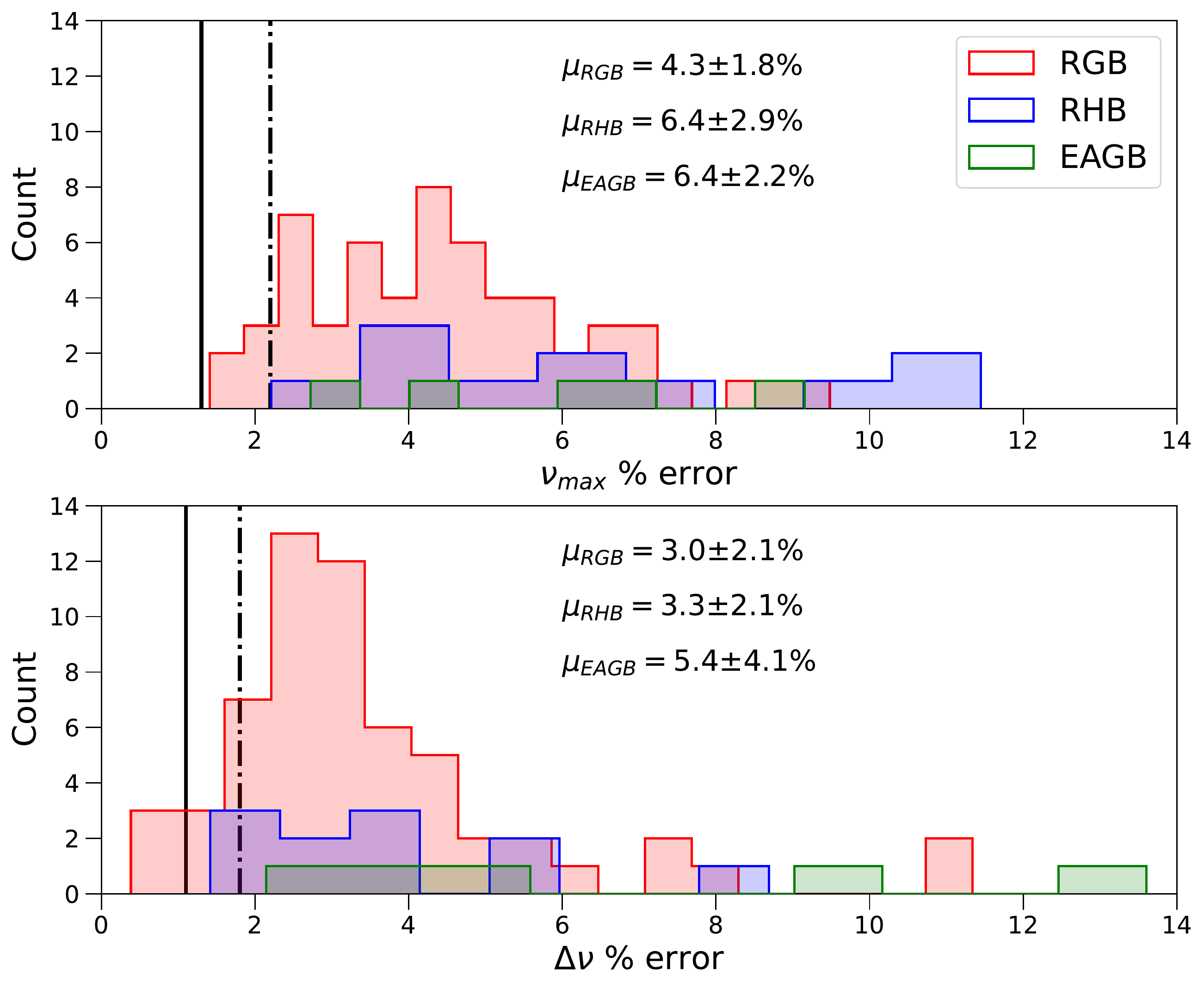}
	\caption{\textbf{a)} Histogram showing the distribution of fractional \% errors from the $\nu_{\text{max}}$ estimations for each evolutionary phase. Our percentage error distributions are compared to the median fractional uncertainties from K2GAP for RGB/EAGB (solid black line) and RHB (dot-dashed black line). Our median fractional uncertainties, $\mu$, and the $1\sigma$ scatter are also written on the plot for each evolutionary phase. \textbf{b)} The same as the top panel but for $\Delta\nu$. }
	\label{fig:percentage_errors}
\end{figure}

We compared our fractional uncertainty distributions for the seismic parameters to the K2 Galactic Archaeology Program data release 3 \citep[K2GAP;][]{K2_GALAH_DR3}, as shown in Fig.~\ref{fig:percentage_errors}. K2GAP did an ensemble study of red giants with \textit{K2} photometry from all campaigns and compared the results from six different global seismology pipelines, including \texttt{SYD}. They reported that the median fractional uncertainty from \textit{K2} data for $\nu_{\text{max}}$ was $1.3\%$ and $\Delta\nu$ was $1.1\%$, in the RGB and EAGB phases. In our study, our median RGB fractional uncertainty and the $1\sigma$ scatter on the value for $\nu_{\text{max}}$ was $4.3\pm1.8\%$, and $3.0\pm2.1\%$  for $\Delta\nu$. These values were even higher for the EAGB stars with the mean fractional uncertainty for $\nu_{\text{max}}$ of $6.4\pm2.2\%$ and $\Delta\nu$ of $5.4\pm4.1\%$. We also found larger fractional uncertainties for the RHB stars with $\nu_{\text{max}}$: $6.4\pm2.9\%$ (this study) vs $2.2\%$ (K2GAP) and $\Delta\nu$: $3.3\pm2.1\%$ (this study) vs $1.8\%$ (K2GAP). This highlights the overall lower SNR (typically $10-15$), and for some stars the lower resolution, of the M4 photometry.

We adopted quality flags based on a visual inspection of the power spectra. Stars observed to have a `noisy' (non-smoothly varying) power excess were labelled as  marginal detections (MD). Otherwise, a detection (D) flag was assigned. The quality flags and global asteroseismic parameters are shown in Table~\ref{tab:final_results}.  

We stress that the low SNR of the power spectra meant that it was especially difficult to measure $\Delta\nu$. We found the majority of our auto-correlation functions did not to show clear repeating patterns, indicating that noise was dominating. The larger fractional uncertainties for EAGB and evolved RGB stars also highlights
the difficulty of measuring $\Delta\nu$ at low frequencies. This is due to the presence of fewer excited modes in solar-like oscillations at lower $\nu_{\text{max}}$. This makes it more challenging to distinguish between signal peaks and noise, especially at the \textit{K2} resolution. This was also reported in the K2GAP data releases \citep{K2GAP_dr1,K2GAP_Dr2,K2_GALAH_DR3}. With such low SNR data, the $\Delta\nu$ values are likely to be more representative of the prior $\Delta\nu$ value used. For these reasons we treat our measured  $\Delta\nu$ values with great caution.


\section{Stellar Parameters}
\label{sec:stellar_params}
\subsection{Effective temperatures \& extinction}
\label{sec:temp&dust}
M4 is located behind the Sco-Oph cloud complex and is affected by substantial reddening. Due to its spatially large size on the sky, it also suffers from significant differential reddening. Studies have suggested that the extinction of light from M4 is due to abnormal dust types \citep[e.g.][]{M4_abnormal_dust1, M4_abnormal_dust2}, hence the standard extinction ratio of $R_V = 3.1$ is inappropriate for this cluster. Using optical and near-infrared photometry to investigate the dust type, \citet[][hereby H12]{Hendricks} determined an extinction ratio of $R_{\text{V}}= 3.62\pm0.07$, where $R_{V} = A_V/E(B-V)$ with $A_V$ being the total extinction in the visual band and $E(B-V)$ being the reddening in the $(B-V)$ colour.

To estimate the line of sight extinction to individual stars, we used the H12 differential dust map for M4. This map can be used to find stars that deviate from the mean extinction of $E(B-V) = 0.37$ mag with a star-to-star scatter of $\sigma = \pm 0.02$ mag. The map contains the centre of the cluster, but doesn't cover our outskirt stars ($\sim$15\% of our sample was positioned outside of the H12 map). For the stars that were within the dust map, we calculated an average extinction and star-to-star scatter of $E(B-V) = 0.36$ mag ($\sigma = \pm 0.02$ mag), slightly lower than the reported average extinction by H12. This is because of the bias in the position of our sample above the cluster centre. The stars not located in the H12 map were assumed to have our average extinction value. 

2MASS $JHK$ magnitudes \citep{2MASS} and $BV$ photometry \citep{BV_photometry, Moch_V_band_mags} were collected for each star and used to calculate the photometric temperatures. \citet{Simon_AGB_paper} showed that calculating the temperature from the $(V-K)$ colour was the most reliable (and had the smallest formal uncertainties) for evolved stars. The $E(B-V)$ values were converted to $E(V-K)$ using Equation 8 in \citet{E(V-K)}. Using the dereddened colours,  $(V-K)_0 = (V-K)-E(V-K)$, temperatures were found using the colour-$T_{\text{eff}}$ relation for giant stars from \citet{Colour_Teff_relation}, with a metallicity of [Fe/H]~$ =-1.1$ \citep{m4_metallicity,M4_metallicity2, chloes_paper}. The largest source of uncertainty in the photometric $T_{\text{eff}}$ were the reddening corrections. 

\begin{figure}
	\centering
	\includegraphics[width=\columnwidth]{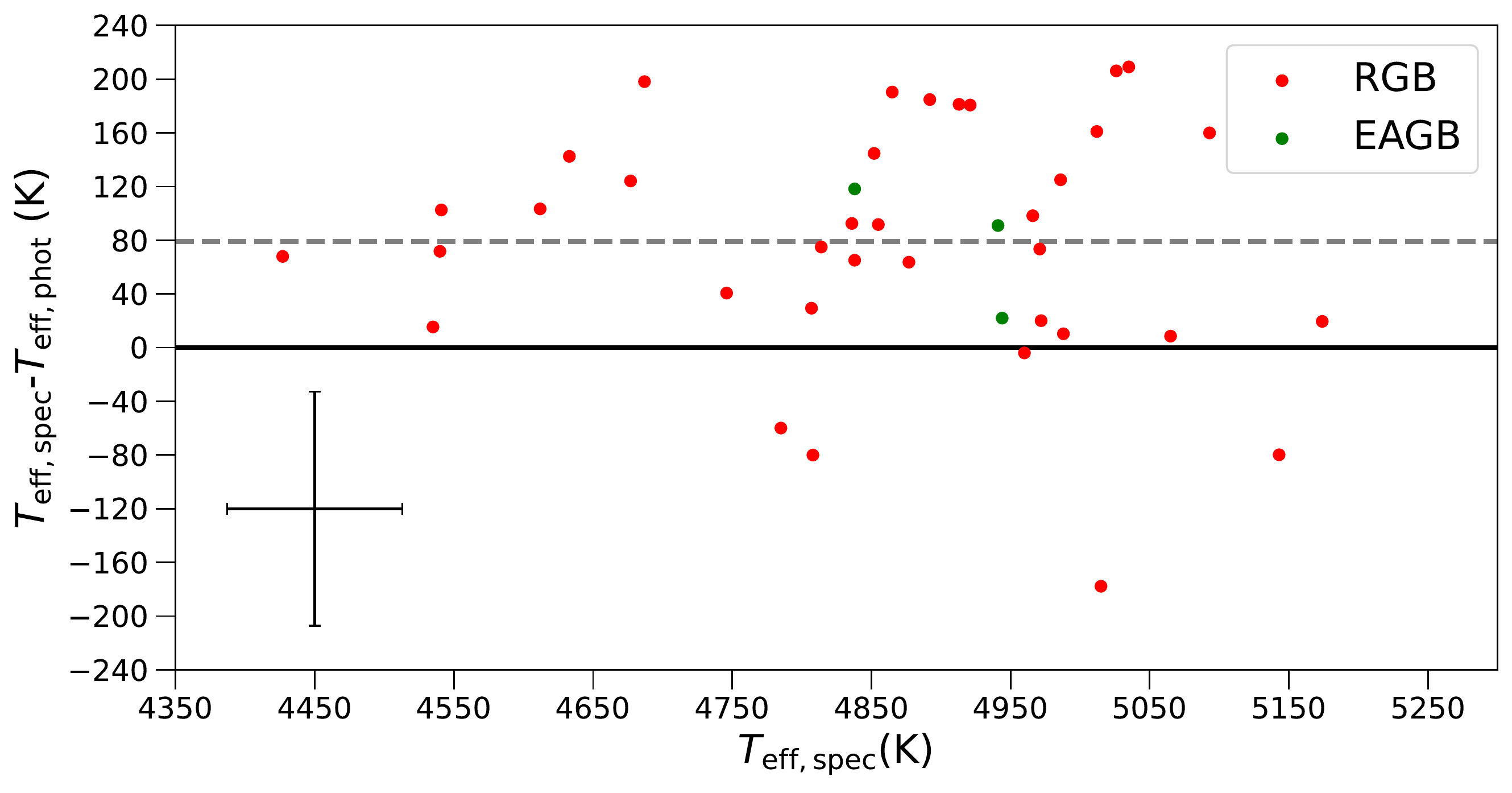}
	\caption{The star-to-star residuals between spectroscopic $T_{\text{eff}}$ from MacLean18 and our photometrically calculated temperatures. The solid line represents zero differences between the two temperature methods. The dashed grey line is the average offset of $81~\mathrm{K}$ between the two methods. A typical error bar is shown (horizontal average spectroscopic uncertainty and vertical is the standard deviation of the differences).}
	\label{fig:teff_offset}
\end{figure}

Ideally, we need an extinction-independent method of calculating $T_{\text{eff}}$, such as spectroscopy. The MacLean18 spectroscopic sample included an overlap of 39 RGB and 3 EAGB stars. An offset of $81~\mathrm{K}$ (Fig.~\ref{fig:teff_offset}) was found between the spectroscopic and photometric $T_{\text{eff}}$, with the spectroscopic temperatures being typically hotter. Since photometric temperatures are dependent on reddening corrections, and often show systematic differences between various colour-$T_{\text{eff}}$ relations (e.g. \citealt{Simon_AGB_paper}), we offset the photometric temperatures by $+81~\mathrm{K}$ for stars without a spectroscopic temperature estimate.  An uncertainty of $108~\mathrm{K}$ was adopted for the scaled temperatures, derived from the addition in quadrature of the average spectroscopic uncertainty and the scatter of the difference between the two temperature methods. Our final $T_{\text{eff}}$ values are provided in Table~\ref{tab:final_results}.

\subsection{Luminosity}
Luminosities were calculated using the bolometric correction ($BC$) for giant stars \citep[Eq. 18 from][]{Bolometric_correction} and the dereddened visual band magnitude $V_0$ using:
\begin{equation}
\label{eq:phot_l}
    \log(L/L_{\odot}) = -0.4\left[V_0-(m-M)_0+BC-M_{\text{bol},\odot}\right]
\end{equation}
For the bolometric magnitude of the Sun, the value of $M_{\text{bol},\odot} = 4.74\pm0.13$ \citep{mbol3, mbol2,mbol_value} was adopted. For the true distance modulus $(m-M)_0$ for M4, we used the average literature value calculated in \citet{dist_mod_M4} of $11.34\pm0.02$. Luminosities can be found in Table~\ref{tab:final_results}.

\subsection{Radii}
\begin{figure}
	\centering
	\includegraphics[width=\columnwidth]{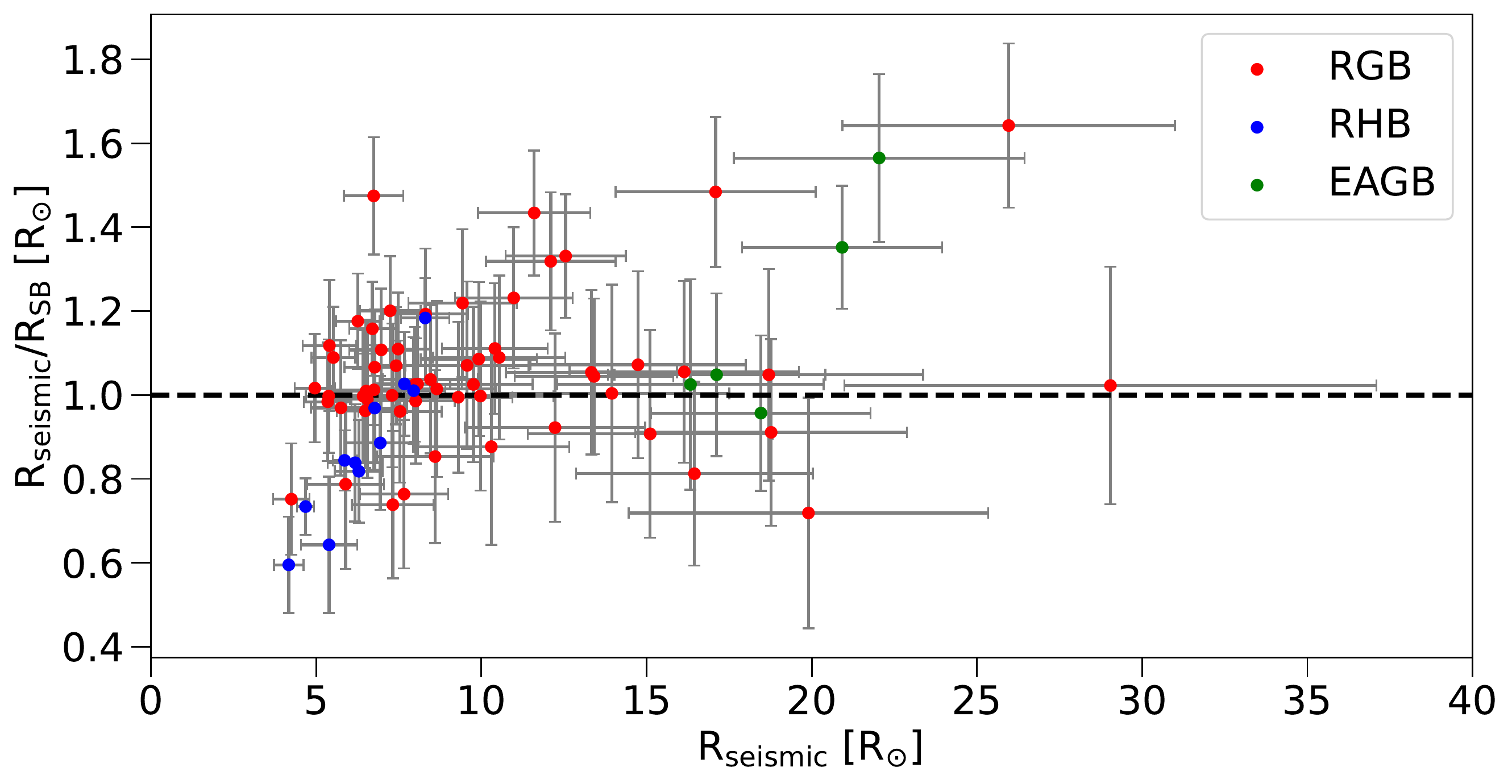}
	\caption{Comparison of the seismic radii and radii calculated from the Stefan-Boltzmann law. Evolutionary status is indicated by the colour coding. The median residual was $2\%$ with a $1\sigma$ scatter of $\pm20\%$. }
	\label{fig:radii}
\end{figure}

We calculated the seismic stellar radius for our sample using:
\begin{equation}
\label{eq:radius_relation}
    \left(\frac{R}{R_{\odot}}\right)\simeq\left(\frac{\nu_{\text{max}}}{\nu_{\text{max},\odot}}\right)\left(\frac{\Delta\nu}{\Delta\nu_{\odot}}\right)^{-2}\left(\frac{T_{\text{eff}}}{T_{\text{eff},\odot}}\right)^{1/2}
\end{equation}
where $\nu_{\mathrm{max,\odot}} =  3090\pm30~\mu\mathrm{Hz}$, $\Delta\nu_{\odot} = 135.1\pm0.1~\mu\mathrm{Hz}$ \citep[][scaled to the \texttt{SYD} pipeline]{dnu_numax_values} and  $T_{\text{eff},\odot} = 5772\pm0.8~\mathrm{K}$ \citep{solar_constants}. These seismic values were compared to independent radius estimates calculated using the Stefan-Boltzmann law (Fig. \ref{fig:radii}). For the ratio of the radius estimates, we found a median residual of $2\%$ with a $1\sigma$ scatter of $\pm20\%$. Seven stars in our sample did not agree within the 2$\sigma$ uncertainties, which could indicate errors in the measurements of the seismic parameters, such as $\Delta\nu$.   

By testing the radii, we can also deduce whether systematics could arise within the asteroseismic scaling relations for certain stars. Zinn et al. (in press) did a comparison study between the seismic radius and radius computed from Gaia parallaxes. They reported that the two types of radius estimates for $R<50~\rsun$\footnote{This is an upward change from the $30~\rsun$ limit in \citet{radii_seismo_systematics}.} agreed within $2\%\pm2\%$(sys), and stars with a radius $R>50~\rsun$ may have systematics in the measurement of the asteroseismic parameters. All of our stars are below this radius limit. The seismic radii estimates are provided in Table~\ref{tab:final_results}.

\section{Mass Results}
\label{sec:mass}
\subsection{Seismic masses \& their averages}
\label{sec:average_mass_results}

Using our estimates for the parameters $\nu_{\text{max}}$, $\Delta\nu$, luminosity, and $T_{\text{eff}}$ (Table~\ref{tab:final_results}), masses of individual stars were determined using the seismic mass equations \ref{eq:mass_relation}-\ref{eq:mass_relation4}. 

As mentioned in Sec.~\ref{sec:intro}, there is some evidence that mass loss begins at around the RGB bump luminosity  \citep{bumpmasslosstheory1,massloss_bump_3,massloss_bump_4}. Since we are interested in measuring the total integrated mass loss between the initial/main-sequence turnoff mass and the HB, including stars above the bump may bias our result. To avoid this problem, we separate the RGB sample into the lower RGB (LRGB; at the luminosity bump and below) and the upper RGB (URGB; above the luminosity bump), where the magnitude of the bump is $V = 13.6$~mag ($G\sim13$~mag; \citealt{M4_RGB_bump_Vmag1}).


It is known that there are deviations from the seismic scaling relation $\Delta\nu \propto \overline{\rho}^{1/2}$, which can introduce systematics into the calculated mass \citep{dnu_corr_theory1,2012NGC67916819, dnu_corr_theory3}. These departures are likely due to differing stellar structures, and it is common to apply corrections to $\Delta\nu$ obtained from models. We implemented $\Delta\nu$ corrections derived from \texttt{asfgrid} \citep{Sharma_asfgrid}. We note that 44\% of our stellar masses were below the lower mass limit of this grid of models, which will result in an under-estimation of the $\Delta\nu$ corrections for these stars. Also, the \texttt{asfgrid} did not have specific $\Delta\nu$ corrections for EAGB stars. We instead treated them as RGB stars to determine their corrections, because they have similar structures.

\begin{figure}
	\centering
	\includegraphics[width=\columnwidth]{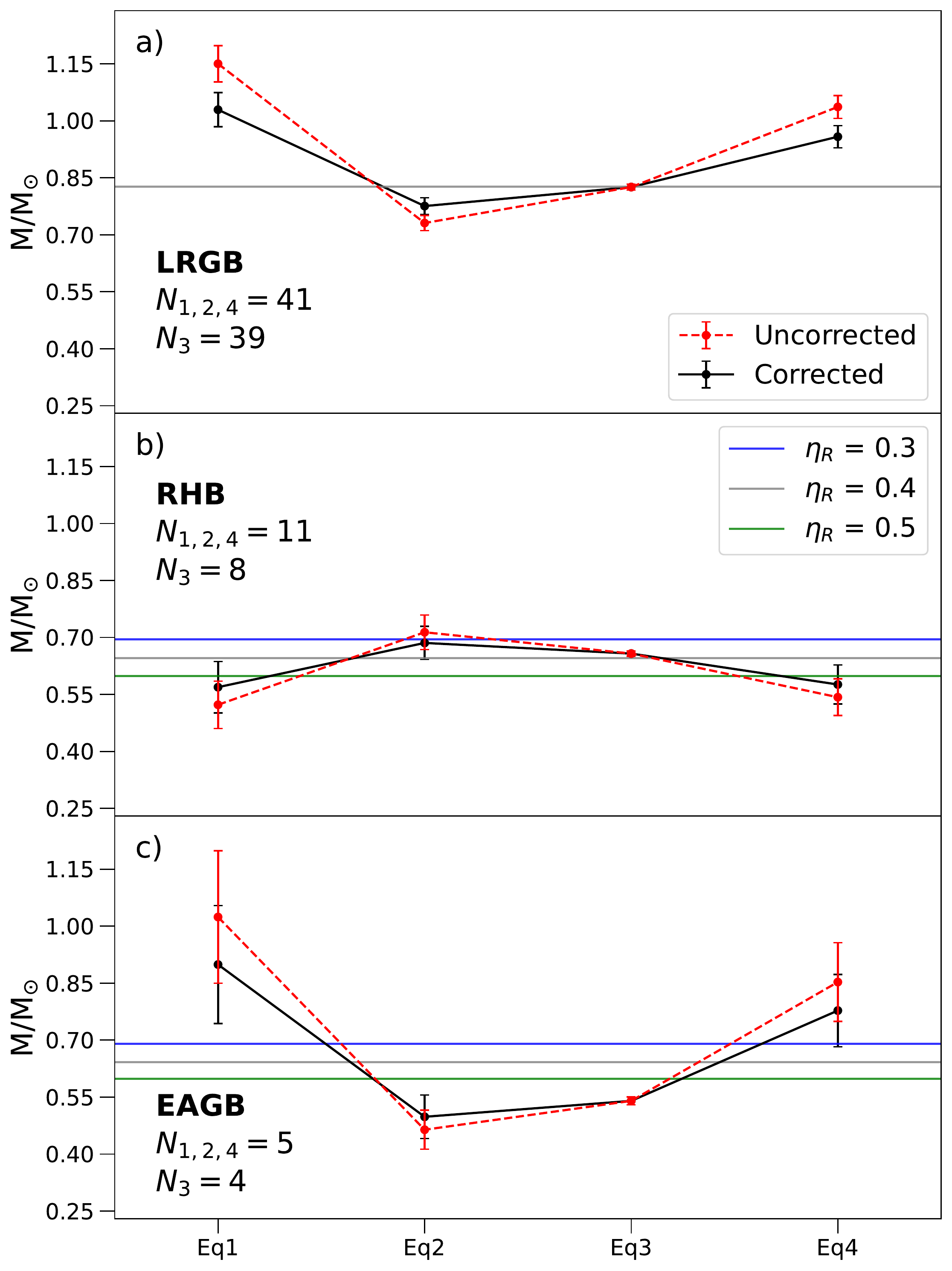}
	\caption{Average masses calculated from the four seismic mass equations for LRGB, RHB, and EAGB for uncorrected (red) and corrected (black; $\Delta\nu$-based Equations 1, 2, and 4). Error bars show the $1\sigma$ standard error on the mean. The solid lines represent model masses as discussed in Sec.~\ref{sec:mass_loss_implications}, where the colours indicate different values of the Reimers' scaling parameter. In the case of the RGB, the solid line shows the average RGB mass which was taken as the initial mass for the models. The number of stars, $N$, used for the average calculations is indicated for each evolutionary phase, where the subscripts represent the equation number. Note that the average mass estimate for Eq.~\ref{eq:mass_relation3} excludes the mass outliers, hence the number of stars, $N_3$, in the mean calculation is less than the other equations (see Sec.~\ref{sec:average_mass_results}).}
	\label{fig:mean_masses}
\end{figure}

We determined average masses for the LRGB, URGB, RHB, and EAGB evolutionary phases using both corrected and uncorrected $\Delta\nu$ mass estimates separately (Table~\ref{tab:mean_masses} and Figure~\ref{fig:mean_masses}). Stars with marginal detections (see Sec.~\ref{sec:seismo} and Table~\ref{tab:final_results}) have been included in these average calculations. We note that without the MD stars our RHB and EAGB samples would be smaller.

Uncertainties for the average masses for each mass relation were taken as the standard error on the mean. 
It can be see from Table~\ref{tab:mean_masses} that the mean masses calculated with Eq.~\ref{eq:mass_relation3} generally have the smallest uncertainties. The Eq.~\ref{eq:mass_relation3} average mass also closely matches the expected masses for the LRGB and RHB from detailed M4 models (MacLean18; \citealt{mcdonald_mass_loss_reimers}; also see Sec.~\ref{sec:mass_loss_implications}). In addition, M16 found a final RGB mass of $0.84~\msun$, in  agreement with our Eq.~\ref{eq:mass_relation3} mean mass of $0.83\pm 0.01~\msun$. This suggests that Eq.~\ref{eq:mass_relation3} may be the most accurate as well as being the most precise.


We found that after $\Delta\nu$ corrections, the masses from Eqs.~\ref{eq:mass_relation}, \ref{eq:mass_relation2} and \ref{eq:mass_relation4} still had larger uncertainties. In Section~\ref{sec:seismo} we highlighted that our $\Delta\nu$ measurements are very unreliable due to the poor SNR of the M4 photometric data. Also, as mentioned above, the $\Delta\nu$ corrections were under-estimated due to \texttt{asfgrid} not covering our mass range for a large fraction of our stars. From Figure~\ref{fig:mean_masses} it is apparent that the $\Delta\nu$ corrections are not sufficient to improve the mass estimates from the $\Delta\nu$-based relations. For these reasons, we only consider our Eq. \ref{eq:mass_relation3} mass estimates as reliable, because they are independent of $\Delta\nu$. Hence, we will only use Eq. \ref{eq:mass_relation3} mass estimates for the following analysis and discussion.

\begin{table*}
\centering
\footnotesize
\caption{\small{The final results for the global seismic quantities, stellar properties, and mass estimates from Eq.~\ref{eq:mass_relation3} for our M4 stars. The star identifications assigned by us (ID) and the Gaia DR2 IDs are provided. The column QF refers to the quality flags, where a `MD' is a marginal detection and `D' is a detection. We also include the Gaia $\mathrm{G}_{\text{mag}}$. Uncertainties for the luminosity, radius, and mass are separated into random and systematic errors, respectively. The unit for the global seismic quantities is $\mathrm{\mu Hz}$, and the $T_{\text{eff}}$ is in $\mathrm{K}$. The full sample will be provided online, and in the supplementary online-only material.} }
\begin{center}
\begin{tabular}{cccccccccc}
\hline
ID       & Gaia DR2 ID         & $\mathrm{G}_{\text{mag}}$ & $\nu_{\text{max}}$ & $\Delta\nu$  & QF & $T_{\text{eff}}$  & $L/L_{\odot}$ & $R/R_{\odot}$ & $M_3/M_{\odot}$ \\ 
\hline
M4AGB01  & 6045466193429562880 & 12.08                                                                 & 9.7$\pm$0.9                                & 1.49$\pm$0.08                       & MD                         & 4944$\pm$29                              & 104.1$\pm$0.9$\pm$12.1                & 23.8$\pm$3.2$\pm$3.2                  & 0.56$\pm$0.05$\pm$0.07                  \\
M4AGB02  & 6045466537026893696 & 11.84                                                                 & 6.7$\pm$0.4                                & 1.39$\pm$0.05                       & D                          & 4838$\pm$62                              & 128.3$\pm$2.6$\pm$14.9                & 18.7$\pm$1.8$\pm$2.8                  & 0.51$\pm$0.04$\pm$0.06                  \\
M4AGB58  & 6045490107808665088 & 11.97                                                                 & 10.3$\pm$0.7                               & 1.76$\pm$0.24                       & D                          & 4840$\pm$108                             & 122.3$\pm$4.1$\pm$14.2                & 17.8$\pm$4.2$\pm$2.7                  & 0.75$\pm$0.08$\pm$0.09                  \\
M4AGB59  & 6045489798570998144 & 11.36                                                                 & 4.9$\pm$0.2                                & 1.16$\pm$0.11                       & MD                         & 5029$\pm$108                             & 209.0$\pm$5.5$\pm$24.3                & 19.8$\pm$3.4$\pm$2.4                  & 0.53$\pm$0.05$\pm$0.06                  \\
M4AGB62  & 6045484266652772864 & 11.93                                                                 & 8.1$\pm$0.2                                & 1.40$\pm$0.03                       & D                          & 4941$\pm$52                              & 125.3$\pm$1.9$\pm$14.5                & 22.6$\pm$1.1$\pm$3.0                  & 0.56$\pm$0.03$\pm$0.07                  \\
M4RGB10  & 6045477837079964928 & 13.40                                                                 & 48.4$\pm$1.0                               & 5.55$\pm$0.14                       & D                          & 4960$\pm$65                              & 32.3$\pm$0.6$\pm$3.7                  & 8.6$\pm$0.5$\pm$1.1                   & 0.86$\pm$0.04$\pm$0.10                  \\
M4RGB104 & 6045477807021629952 & 13.15                                                                 & 34.1$\pm$1.9                               & 4.20$\pm$0.05                       & D                          & 4851$\pm$108                             & 38.8$\pm$1.3$\pm$4.5                  & 10.4$\pm$0.6$\pm$1.6                  & 0.79$\pm$0.08$\pm$0.09                  \\
M4RGB11  & 6045477905795000448 & 12.22                                                                 & 10.7$\pm$0.5                               & 1.77$\pm$0.05                       & D                          & 4612$\pm$58                              & 92.9$\pm$2.4$\pm$10.8                 & 18.1$\pm$1.3$\pm$3.4                  & 0.71$\pm$0.05$\pm$0.08                  \\
M4RGB124 & 6045466124710020608 & 14.05                                                                 & 91.4$\pm$3.3                               & 9.41$\pm$0.26                       & D                          & 5088$\pm$108                             & 17.1$\pm$0.4$\pm$2.0                  & 5.7$\pm$0.4$\pm$0.6                   & 0.79$\pm$0.07$\pm$0.09                  \\
... &&&&&&&&&\\
\hline
\label{tab:final_results}
\end{tabular}
\end{center}
\end{table*}

\begin{table*}
\footnotesize
\caption{Average masses, $\overline{M}$, calculated for each evolutionary phase from Eqs~ \ref{eq:mass_relation}-\ref{eq:mass_relation4}. The subscript \textit{corr} indicates the mean mass estimates that used the $\Delta\nu$-corrected masses. An extra mass column is shown for Eq~\ref{eq:mass_relation3}, where we excluded mass outliers from the average mass ($\overline{M}_{3,\text{without outliers}}$). The uncertainty is the standard error on the mean. All units in terms of $\msun$. }
    \begin{tabular}{ccccccccc}
      \hline 
      Evol Stage & $\overline{M}_1$ & $\overline{M}_{1,\text{corr}}$ & $\overline{M}_2$ & $\overline{M}_{2,\text{corr}}$ & $\overline{M}_{3}$ & $\overline{M}_{3,\text{without outliers}}$ & $\overline{M}_4$ & $\overline{M}_{4,\text{corr}}$ \\ 
      \hline 
      LRGB & $1.15\pm0.05$ & $1.03\pm0.04$ & $0.73\pm0.02$ & $0.78\pm0.02$ & $0.83\pm0.01$ & $0.826\pm0.008$ & $1.04\pm0.03$ & $0.96\pm0.03$\\
      URGB & $1.01\pm0.09$ & $0.83\pm0.07$ & $0.68\pm0.03$ & $0.75\pm0.04$ & $0.75\pm0.01$ & $0.75\pm0.01$ & $0.91\pm0.05$ & $0.79\pm0.05$ \\
      RHB & $0.52\pm0.06$ & $0.57\pm0.07$ & $0.71\pm0.05$ & $0.69\pm0.04$ & $0.62\pm0.02$ & $0.658\pm0.006$ & $0.54\pm0.05$ & $0.58\pm0.05$ \\
      EAGB & $1.02\pm0.17$ & $0.90\pm0.16$ & $0.46\pm0.03$ & $0.50\pm0.06$ & $0.58\pm0.04$ & $0.54\pm0.01$ & $0.84\pm0.13$ & $0.78\pm0.10$\\
      \hline 
    \end{tabular}
\label{tab:mean_masses}
\end{table*}

\begin{figure*}
	\centering
	\includegraphics[width=\textwidth]{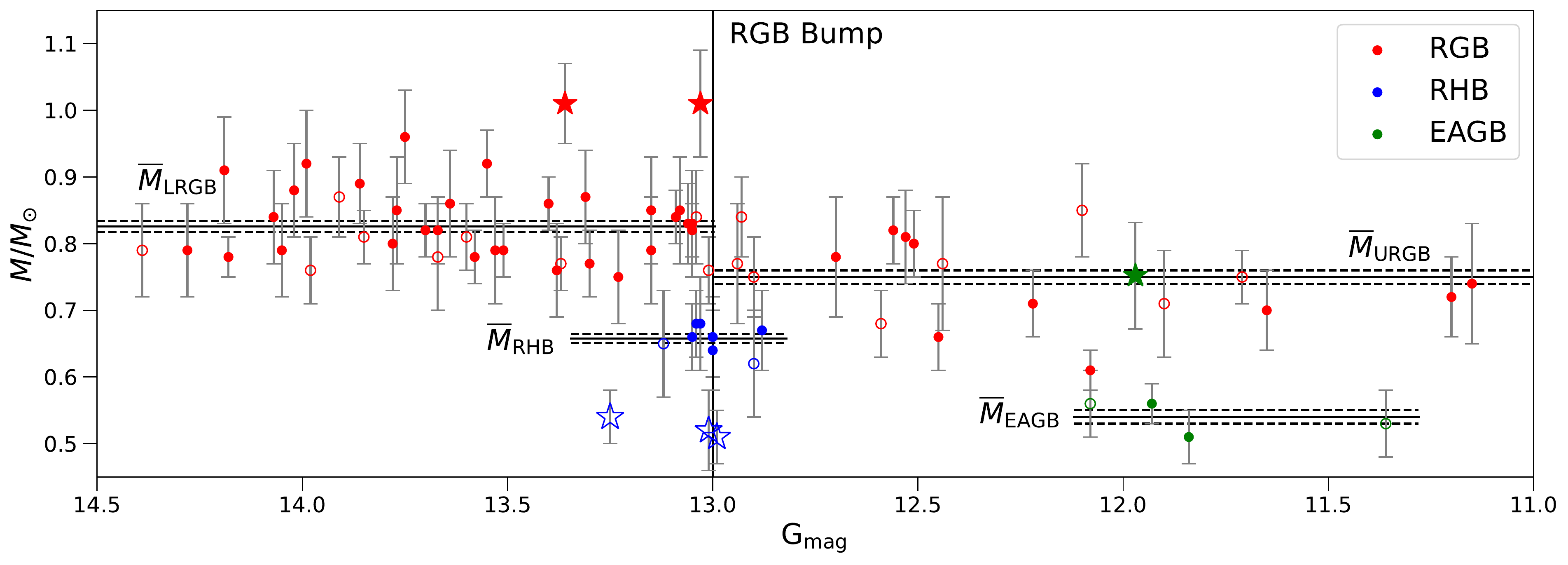}
	\caption{Masses for our entire sample using Eq.~\ref{eq:mass_relation3}. Open symbols indicate stars with marginal detections (Sec.~\ref{sec:seismo}). The random uncertainties for the mass estimates are represented by the error bars. The mean masses (without outliers; see Sec.~\ref{sec:average_mass_results}) for the RGB (split by the RGB bump magnitude represented by the black solid line), RHB, and EAGB, are shown by the black solid horizontal lines across the magnitude range for each evolutionary phase, and appropriately labelled. The 1$\sigma$ uncertainties on the means are indicated by the black dashed lines. Mass outliers are shown by star symbols.}
	\label{fig:masses_eq3}
\end{figure*}

Figure~\ref{fig:masses_eq3} shows the individual masses (Eq.~\ref{eq:mass_relation3}) for our entire sample. Six stars (indicated by star symbols) were identified as obvious mass outliers (approximately $2\sigma$ from the average mass in each evolutionary phase): two over-massive RGB stars, one over-massive EAGB star, and three under-massive RHB stars. We discuss the possible causes for the outliers in Sections~\ref{sec:mass_outliers} and \ref{sec:non_standard_evol}. The outliers were removed from the mean mass calculation, to avoid the averages being biased. Our final mean mass estimates were $0.826\pm0.008~\msun$ (LRGB), $0.658\pm0.006~\msun$ (RHB), and $0.54\pm0.01~\msun$ (EAGB). The final mean mass estimates are provided in Table~\ref{tab:mean_masses} (denoted by the `without outliers' subscript) and shown in Figure~\ref{fig:mean_masses}. We identified that the EAGB mean mass is significantly lower than predicted model estimates (Fig.~\ref{fig:mean_masses}c), which will be discussed further in Section~\ref{sec:EAGB_masses}.

We compared our mean mass estimates to the Eq.~\ref{eq:mass_relation3} mean mass results of the M16 study. Their RGB mean mass of $0.84\pm0.01~\msun$ is consistent with ours within the 2$\sigma$ uncertainties. However, the M16 RHB mass of $0.61\pm0.08~\msun$ is outside the 2$\sigma$ uncertainties for our RHB mean mass. This could be because the RHB mass estimate from M16 was determined using only a single star, compared to our sample of 11 stars. In Sec.~\ref{sec:mass_loss_implications}, we compare our masses to other studies that used independent mass determination methods.

\subsection{Mass loss results}
\label{sec:mass_loss_results}
Total integrated mass loss on the RGB was determined by taking the difference between the average mass of the LRGB and RHB phases of evolution, denoted by $\Delta\overline{M}$. The standard error on the means for the LRGB and RHB were added in quadrature to determine the associated uncertainties for the mass differences. These mass difference uncertainties were of the same order of magnitude as the \citet{2012NGC67916819} open cluster mass loss study, which used a different error propagation method. We found an integrated mass loss on the RGB of $\Delta\overline{M}=0.17\pm0.01~\msun$. This value is similar to the expected total integrated RGB mass loss $\sim0.2~\msun$ for GCs (e.g. \citealt{Macdonald_omega_centauri_massloss,tucan_47}; also see Sec.~\ref{sec:mass_loss_implications}). 




Upon inspection of the 18 URGB stars, we saw evidence of a decline in masses (Table~\ref{tab:mean_masses} and Fig.~\ref{fig:masses_eq3}), which supports the theory that mass loss starts becoming significant at the RGB bump. We also saw that a substantial amount of matter is lost straight after the bump. This is in contradiction to the Reimers' and Schr\"oder \& Cuntz schemes, where the mass loss rates slowly increase on the RGB until the majority of the mass is lost at the RGB tip. We further explore this observation in detail in an upcoming paper.  

\subsection{Mass outliers}
\label{sec:mass_outliers}

In Fig.~\ref{fig:masses_eq3} the mass outlier stars are displayed as large star symbols. Here we consider if systematic uncertainties can explain the deviant masses. In Sec.~\ref{sec:non_standard_evol} we discuss the possible interpretations if these masses are correct.

\subsubsection{Over-Massive Cluster Members: M4RGB36, M4RGB217 \& M4AGB58}
\label{sec:Massive}
Three stars were found to have masses significantly larger than the average value for their evolutionary phase; $M_{\text{M4RGB36}}=1.01\pm0.08(\text{random})\pm0.12(\text{sys})~\msun$, $M_{\text{M4RGB217}}=1.01\pm0.06(\text{random})\pm0.12(\text{sys})~\msun$, and $M_{\text{M4AGB58}}=0.75\pm0.08(\text{random})\pm0.09(\text{sys})~\msun$. These stars have been identified as members from proper motions and parallaxes, and have good asteroseismic parameter measurements (all stars have the `D' quality flag). We considered that the larger masses are due to systematics in $T_{\text{eff}}$ or luminosity, where differential reddening introduces the greatest uncertainty.

\begin{figure}
	\centering
	\includegraphics[width=\columnwidth]{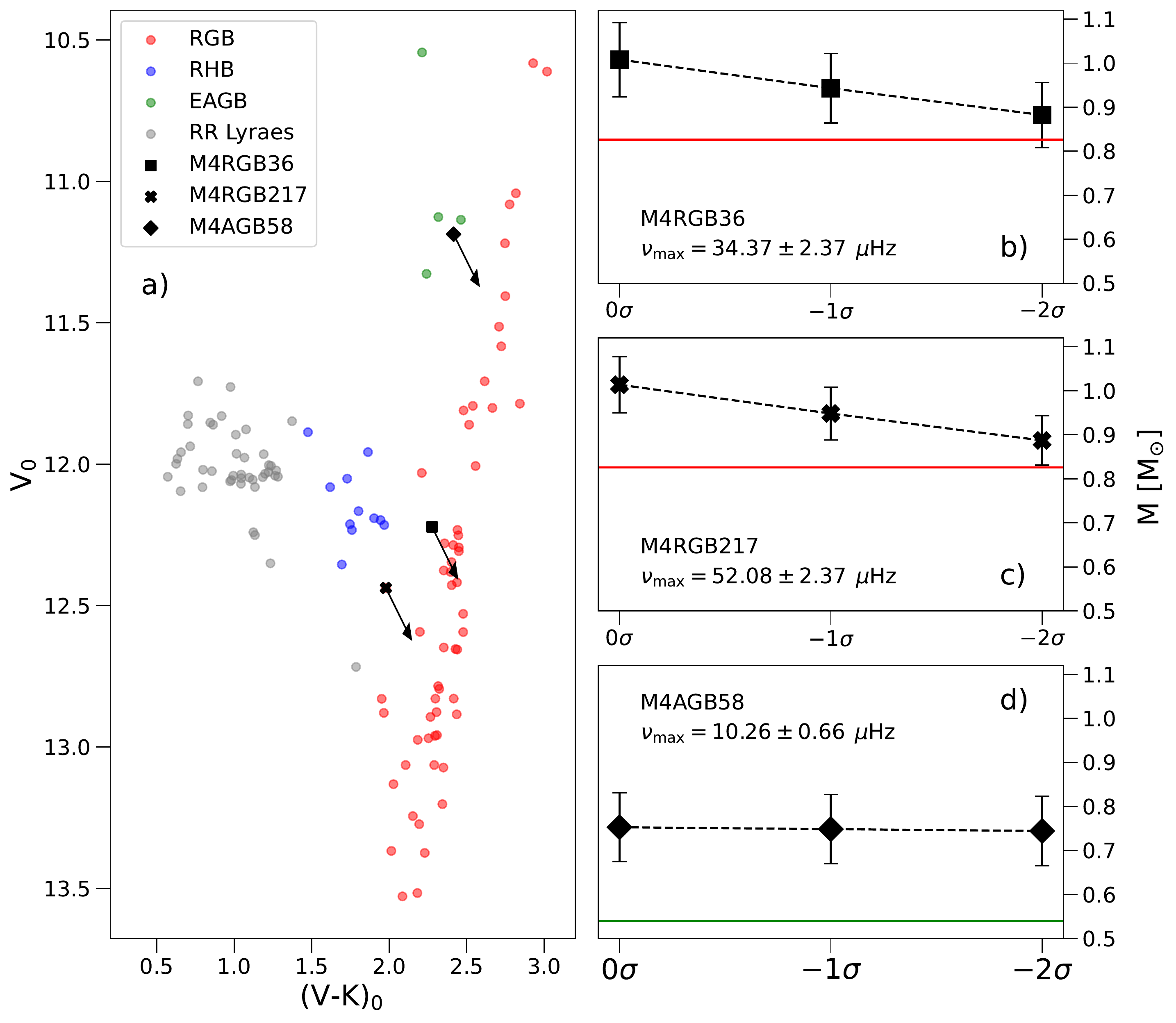}
	\caption{\textbf{a)} Reddening-corrected CMD in the V magnitude and $(V-K)$ colour. The position in the CMD for each over-massive star with the original extinction is represented by the black symbols. The arrows indicate the maximum change in the magnitude and colour when decreasing the reddening correction by $2\sigma$ (see text for details). \textbf{b,c,d)} The calculated masses for the original extinction correction ($0\sigma$) and when subtracting $1\sigma$ and $2\sigma$ scatter in the reddening corrections. Each panel is labelled with the star ID and measured $\nu_{\text{max}}$. Solid horizontal lines indicate the calculated average mass for the corresponding evolutionary phases.}
	\label{fig:dust_tests}
\end{figure}

These stars were contained within the H12 map and were all found to have an extinction correction of $E(B-V)=0.37$ mag. We note that there are difficulties when mapping the differential extinction in M4, such as the large pixel size ($20$''/pixel) in the H12 dust map, which can mask local variations. Therefore, the dust extinction of these stars may be different than that reported by the H12 map. To test the sensitivity of our mass estimates to reddening corrections, we perturb our $E(B-V)$ values by subtracting the $1\sigma$ and $2\sigma$ dust scatter ($\sigma = 0.02$ from Section~\ref{sec:temp&dust}). The effects of these extinction tests on the over-massive stars' photometry and mass estimates are shown in Fig. \ref{fig:dust_tests}.   

M4RGB36 and M4RGB217 had spectroscopic temperatures from MacLean18, which meant that the dust correction dependency was only in the luminosity. By inspection of the offset between the spectroscopic and photometric temperatures for these stars, we noticed that they deviated significantly from the average offset of $81\mathrm{K}$, further suggesting that a wrong reddening correction has been used. When reducing the reddening correction for both of these stars, their position in the CMD moved closer to the centre of the RGB. Their masses also decreased by $0.06-0.07~\msun$ each step, to within 1.5 standard deviations of the mean for the LRGB. Hence, M4RGB36 and M4GB217 would not be identified as outliers if the reddening correction were 2$\sigma$ less than that determined from the H12 map. 

The EAGB star, M4AGB58, was 40\% more massive than the mean mass for this evolutionary phase. This mass is similar to the URGB masses at the same Gaia magnitude, indicating that it could be a misclassified RGB star. However, after reducing the reddening corrections, this star would have an ambiguous classification because it lies between the EAGB at $-1\sigma$ and a cautious RGB classification at $-2\sigma$. This star did not have a spectroscopic temperature and thus the calculated photometric temperature was used. However, these changes in the reddening corrections had little effect on the final mass of the star (Figure \ref{fig:dust_tests}), where there was a mass difference of $\sim 0.004~\msun$ for each $\sigma$ calculation. This is very different to what was seen for the massive RGB stars. This is due to different reddening dependencies in  $T_{\text{eff}}$ and luminosity, which counteract each other. A spectroscopic $T_{\text{eff}}$ for M4AGB58 is needed to investigate how the mass changes with dust corrections. 

These results indicate that the larger masses for the three stars could be due to inaccurate dust corrections. However, without a finer resolution dust map of M4, this cannot be confirmed.

\subsubsection{Under-massive stars: M4RHB47, M4RHB127, \& M4RHB246}
\label{sec:Undermassive}
Our sample also contained three apparently under-massive stars in the RHB stage with masses; $M_{\text{M4RHB47}} =0.52\pm0.06(\text{random})\pm0.06(\text{sys}))~\msun$, $M_{\text{M4RHB127}} = 0.51\pm0.04(\text{random})\pm0.06(\text{sys}))~\msun$, and $M_{\text{M4RHB246}} =0.54\pm0.04(\text{random})\pm0.06(\text{sys})~\msun$.  These masses are lower than the theorised sub-population-2 HB mass of $0.6~\msun$ (see Section~\ref{sec:multi_pops}). All three under-massive stars were identified as marginal dectections, implying a poor seismic parameter measurement. Better photometry is needed to improve the seismic result, although it would require a shift of $\sim9~\mu\mathrm{Hz}$ for these stars to have a mass equal to the mean RHB mass from this study. Such a shift is incompatible with the data to displace $\nu_{\text{max}}$ by this amount (typical RHB $\nu_{\text{max}}$ uncertainty is $\pm2~\mu\mathrm{Hz}$). 

\begin{figure}
	\centering
	\includegraphics[width=\columnwidth]{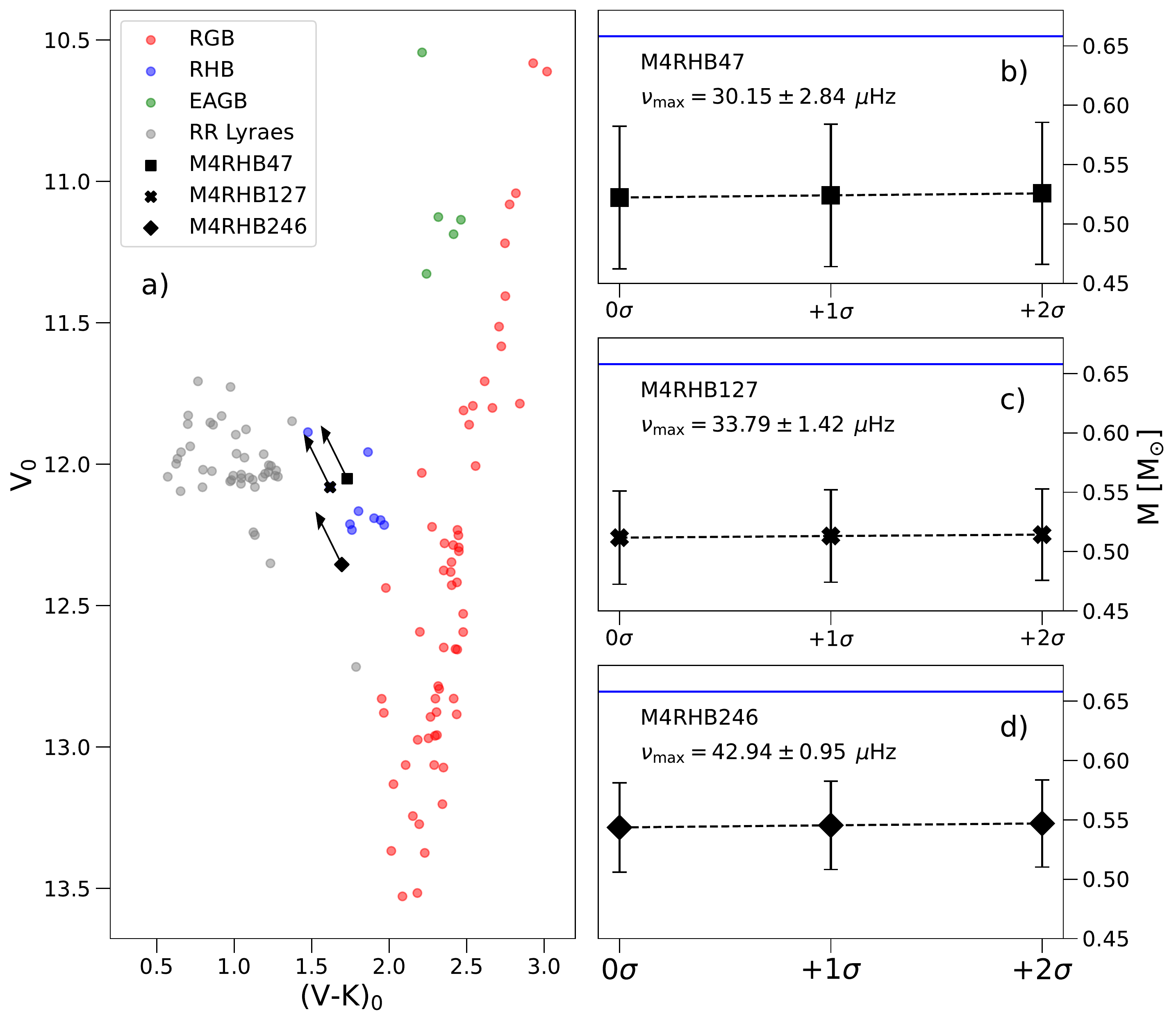}
	\caption{Same as Figure \ref{fig:dust_tests} but for the apparently under-massive stars, and an addition of the $1\sigma$ and $2\sigma$ scatter to the reddening corrections. Blue horizontal lines in panels a,b, and c indicate the calculated mean mass for this phase of evolution. }
	\label{fig:dust_tests_undermassive}
\end{figure}

We again considered inaccurate reddening corrections for the unexpected mass estimates. A similar dust test to the over-massive stars was made, but instead of subtracting by $1\sigma$ and $2\sigma$ to the reddening correction, we added it (Fig.~\ref{fig:dust_tests_undermassive}). The original H12 reddening corrections for M4RHB47 was $E(B-V) = 0.37$ and $E(B-V) = 0.36$ for the stars M4RHB127 and M4RGB246. Our change to the reddening made the stars hotter and more luminous. Photometric temperatures were used for these three RHB stars, and hence the true effect of the reddening correction on the mass cannot be determined (as seen for the massive star M4AGB58 in Sec.~\ref{sec:Massive}) -- again, spectroscopic temperatures would be very useful for resolving the issues caused by reddening. 

\section{Discussion}
\label{sec:discussion}
\subsection{Implications for stellar models and comparison to other methods for determining mass loss}
\label{sec:mass_loss_implications}
Our results for the average LRGB mass, $0.826\pm0.008~\msun$, and average RHB mass, $0.658\pm0.006~\msun$ (Table~\ref{tab:mean_masses}), can be used as constraints for mass loss in stellar models. Here we make the assumption that the initial masses of the LRGB and RHB stars are equal, because the mass difference ($\sim 10^{-3}~\msun$) is smaller than our mass uncertainties (see Sec.~\ref{sec:intro}).
In particular, the masses can be used to calibrate mass-loss scaling parameters, such as $\eta_{\text{R}}$ in the most commonly used \cite{Reimers_massloss_rate} formula. To explore this we calculated a series of models using the MESA stellar code (\citealt{paxton2011,paxton2019}; version 12778). The models all have identical initial parameters and input physics except that $\eta_{\text{R}}$ is varied. The initial mass of the models was taken as our measured average mass for the M4 LRGB stars, and the initial metallicity as $Z = 0.0015$, based on the measured [Fe/H]$=-1.15$ (\citealt{chloes_paper}) plus an alpha-enhancement. The initial helium abundance was taken as Y~$=0.245$ to match that for the first subpopulation of M4, which populates the red end of the HB (\citealt{Marino_HBs,chloes_paper}). We used the standard MESA nuclear network (`basic.net'), standard equation of state (see \citealt{paxton2019} for details), and mixing length parameter $\alpha = 2.0$. Convective boundary locations were based on the Schwarzschild criterion, extended with exponential overshoot (\citealt{herwig1997}; $f_{os} = 0.016$) at the bottom of the RGB envelope and at the top of the convective core during core helium burning.

We found a model value of $\eta_{\text{R}} = 0.39$ closely matched our measured integrated RGB mass loss, giving an RHB mass of 0.657~M$_{\odot}$, compared to our measured average RHB mass of 0.658~M$_{\odot}$. The mass loss rate \textit{along} the HB itself is unknown, however a commonly used assumption is to extend the Reimers mass loss formalism to the HB. At the lower luminosity of the HB (compared to the upper RGB), there is expected to be little mass loss. For our $\eta_{\text{R}} = 0.39$ model we find a total mass loss of $0.01~\msun$ using this assumption. This is smaller than our mass uncertainties on individual HB stars, and of a similar magnitude to the $1\sigma$ scatter of $0.02~\msun$ among our HB masses.

To illustrate the sensitivity of modelled HB mass to $\eta_{\text{R}}$, we show the RHB masses for three values of $\eta_{\text{R}} = 0.30, 0.40, 0.50$ in Figure~\ref{fig:mean_masses}b. Taking Eq.~\ref{eq:mass_relation3} as the most reliable (as discussed earlier) it can be seen that $\eta_{\text{R}} = 0.30$ gives an HB mass that is too high (0.70~M$_{\odot}$) compared to our observed average mass of 0.658~$\msun$, and $\eta_{\text{R}} = 0.50$ gives mass that is to low (0.60~M$_{\odot}$). However, the observed average RHB mass is reasonably well matched by $\eta_{\text{R}} = 0.40$, which has an RHB mass of 0.65~M$_{\odot}$ As mentioned above, our $\eta_{\text{R}} = 0.39$ model gives an almost exact match, however given the inherent uncertainties in stellar models we believe it unwise to place too much credence at this level of precision.

Previous studies have estimated $\eta_{\text{R}}$ for reproducing the observed characteristics of M4 RHB stars using independent methods. By comparing a grid of models with observed CMDs, \cite{mcdonald_mass_loss_reimers} found $\eta_{\text{R}} = 0.40^{+0.05}_{-0.06}$. This matches with our optimal value of 0.39. \cite{chloes_paper} used a similar method and found $\eta_{\text{R}} = 0.44$, consistent (within uncertainties) with the value of \cite{mcdonald_mass_loss_reimers}, and slightly higher than our optimal value.

Although the $\eta_{\text{R}}$ values all agree, this could be a coincidence, since the value is degenerate with different initial masses and HB masses. However this doesn't appear to be the case with M4, since previous studies also agree on the initial mass and HB mass:  \cite{mcdonald_mass_loss_reimers} found an initial mass of $0.84^{+0.02}_{-0.02}$, consistent with our LRGB mass. Likewise, \cite{chloes_paper} found an initial mass of 0.827~M$_{\odot}$, almost identical to our value of 0.826~M$_{\odot}$. For their average HB mass \cite{mcdonald_mass_loss_reimers} report $0.66\pm0.01~\msun$, also in agreement with our value of $0.658 \pm 0.006~\msun$. \cite{chloes_paper} did not report a HB mass.
%
%

\cite{infrared} determined an empirical mass-loss relation for Pop. II giants based on mid-IR photometry of RGB stars. Their Equation~4, which is dependent only on [Fe/H], enables us to calculate the total expected RGB mass loss for M4, giving $0.15\pm 0.03~\msun$. This matches with our integrated mass-loss measurement of $\Delta\overline{M}=0.17\pm0.01~\msun$.

The remarkable agreement of initial mass, HB mass, $\eta_{\text{R}}$, and total RGB mass loss between these four independent studies, and three very different methods -- CMDs+models, IR observations, and asteroseismology -- is very reassuring. It confirms that the traditional methods are accurate. Asteroseismology does however give higher precision -- even with the low-quality K2 data used in the current study.

Turning to the next phase of stellar evolution, in Fig.~\ref{fig:mean_masses}c we include the model EAGB masses for the same set of model $\eta_{\text{R}}$ values. In this case, \textit{none of the models can reproduce the average EAGB mass}. This surprising mismatch could have a number of different explanations, which we discuss in Sec.~\ref{sec:EAGB_masses}.

\subsection{Multiple populations: bi-modal mass distribution?}
\label{sec:multi_pops}
It is well established that GCs contain multiple populations defined by their light elemental abundances \citep[e.g. C, N, O, Na, O, and He;][]{light_elemental_abundances1,light_elemental_abundances2}. Among these abundance variations, helium is a key element for stellar evolution: a star that has a higher He abundance will evolve faster \citep{chloes_paper,2pops_models}. Because GC stellar populations are essentially co-eval, this means that the multiple populations will have different current masses -- the more `massive' He-poor stars and the lower-mass He-rich stars. 

The two populations in M4 have been observed and modelled by multiple studies \citep[e.g.][]{M4_multi_pops1,M4_multi_pops2,M4_multi_pops3, chloes_paper,massloss_2pops,M4_multi_pops4}. In MacLean18's two population models, the first sub-population (SP1; lower light metal abundances and more massive) and the second sub-population (SP2; higher light metal abundances and less massive) had an initial mass difference of $0.04~\msun$. This mass difference is of a similar magnitude to our individual mass uncertainties (Table~\ref{tab:final_results}), and hence could be too small to detect in the RGB sample. 

\begin{figure*}
	\centering
	\includegraphics[width=0.8\textwidth]{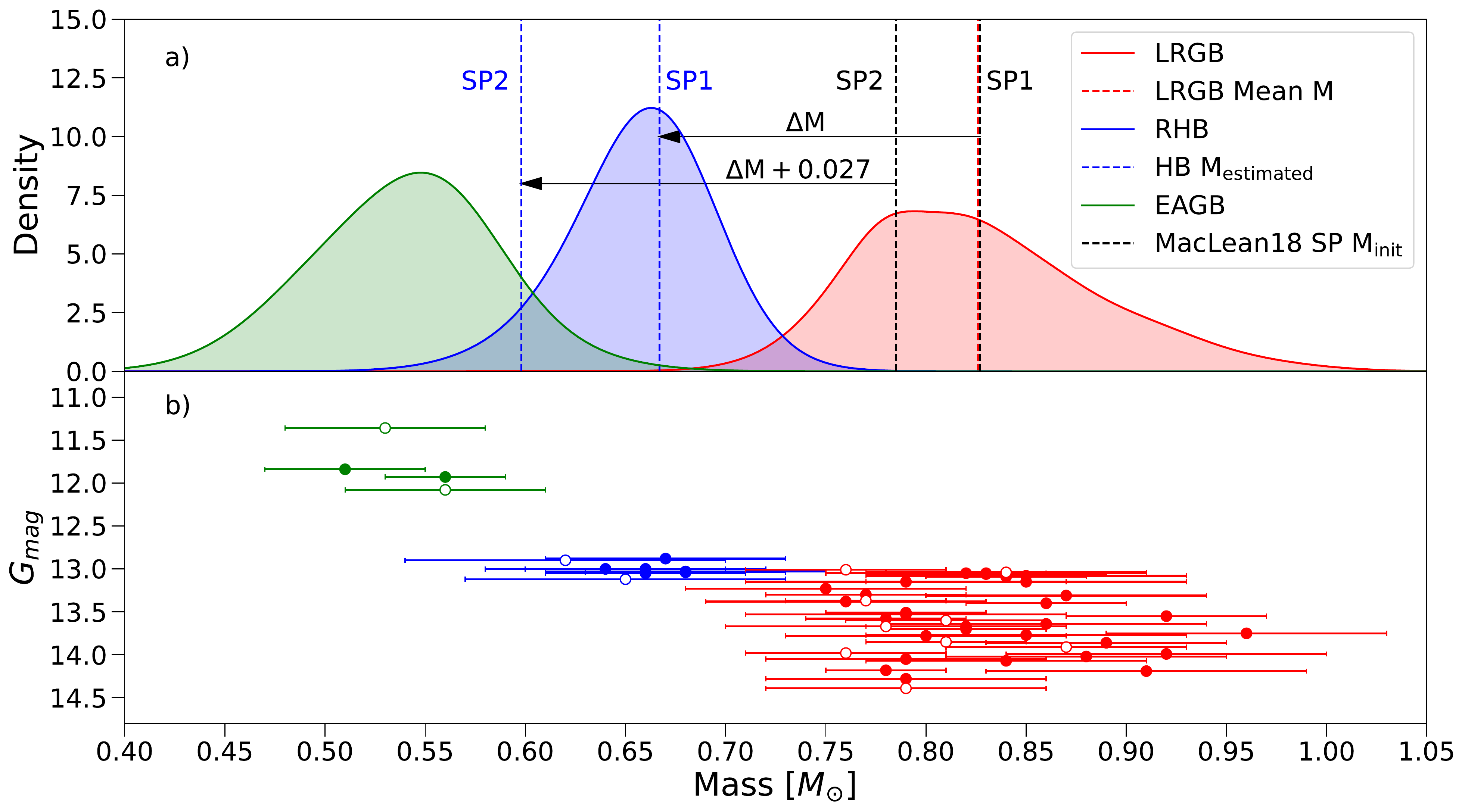}
	\caption{\textbf{a)} Gaussian kernel density estimation (KDE) histograms showing the seismic mass distributions for the LRGB, RHB, and EAGB samples (without outliers). Stars with marginal detections are shown as open circles. The $1\sigma$ random uncertainties on the mass estimates were used to define the histogram bin widths for the RHB and LRGB. Due to the small EAGB sample, $2\sigma$ uncertainties were used. We illustrate the expected masses of the sub-populations by vertical lines; the initial masses for both M4 sub-populations from MacLean18 (black) and the HB masses (blue, see Sec.~\ref{sec:multi_pops}). Note that the MacLean18 SP1 mass overlies our LRGB mean mass (0.827 vs 0.826~$\msun$, respectively). Arrows indicate the RGB mass loss adopted, where $\Delta\mathrm{M}$ is the RGB mass difference calculated from Eq~\ref{eq:mass_relation3}. The HB SP2 stars are expected to be on the blue HB. The mean LRGB mass is also shown by the red line.  \textbf{b)} Mass estimates with random uncertainties against Gaia magnitude.}
	\label{fig:KDE1}
\end{figure*}

To test this, we plotted the mass distributions for the LRGB, RHB, and EAGB (without outliers) in kernel density estimation (KDE) histograms (Fig. \ref{fig:KDE1}a). We used the random uncertainties of the masses for the Gaussian widths. We find a broad mass distribution on the LRGB, larger than expected for a single population. In Fig. \ref{fig:KDE1}a we over-plot the MacLean18 initial masses for their sub-population models. Although our mass distribution is not clearly bi-modal, the MacLean18 SP1 and SP2 initial masses fall at either end of the broad peak. This suggests that this distribution could be due to the two sub-populations, but is smoothed by our (relatively) large uncertainties.

Our RHB mass distribution is consistent with the presence of only one sub-population. This concurs with spectroscopic studies that generally show that the RHB samples in GCs only contain first sub-population stars, and the blue HBs only contain second population (or higher) stars. For M4, \cite{Marino_HBs} showed that the RHB is populated only by Na-poor (SP1) stars. In Fig.~\ref{fig:KDE1} we also show the observed SP1 and predicted SP2 HB sub-population masses, where SP1 experiences our observed integrated RGB mass loss of $0.17 \pm 0.01~\msun$ (Sec.~\ref{sec:mass_loss_results}), and SP2 has an adopted mass loss that is $0.027~\msun$ larger than SP1 (as calculated by \citealt{massloss_2pops}). As can be seen, there is no peak in our distributions near the predicted HB SP2 mass, as expected.

Interestingly, we do not detect a bi-modality in our EAGB mass distribution. This is in agreement with the observations by MacLean18, who report that M4 lacks SP2 stars at this evolutionary phase (although there is some disagreement in the literature on this, see \citealt{Marino2017}).

To extend our investigations into the multiple populations of M4, we require chemical information for sub-population membership. Despite the limitations (low SNR of the \textit{K2} data, photometric $T_{\text{eff}}$, and small sample size), our study hints at what could be accomplished with the synergy between asteroseismology and spectroscopy in studying the multiple populations of globular clusters.

\subsection{EAGB masses}
\label{sec:EAGB_masses}
From our 5 EAGB stars, we measured a mean mass of $0.54\pm0.01~\msun$, significantly lower than the modelled EAGB mass of $0.64~\msun$ for $\eta_R = 0.40$ (Fig.~\ref{fig:mean_masses}). For our measured RHB mean mass of $0.658\pm0.006~\msun$, the lower mass of the EAGB suggests there is a significant mass loss between the helium burning stage and the EAGB of $\sim0.12~\msun$. Mass loss on the HB in low mass stars has been modelled, but normally in the context of explaining HB morphology, such as the extreme HB (stars hotter than the blue HB; e.g. \citealt{Yong_HB_massloss,Vink_HB_massloss}). There are no detailed studies that consider the mass loss of cooler RHB stars, and it is typically assumed there is either very little ($\sim0.01~\msun$, see Sec.~\ref{sec:mass_loss_implications}), or no mass loss during this phase.


Interestingly, our EAGB mean mass estimate is consistent with the measured white dwarf mass for GCs of $0.5-0.55~\msun$ \citep{GC_WD_masses}. For low mass stars, the AGB phase of evolution can be separated into two parts; the EAGB and the thermal-pulsing AGB stage, where intermittent helium shell flashes occur. After the thermal-pulsing AGB stage, a star then evolves to become a white dwarf. If our low EAGB masses are correct, it would suggest that these stars do not reach the thermal-pulsing AGB stage, and become white dwarfs earlier in their evolution. These objects are similar to the theorised AGB-manqu\'e stars (\citealt{HB_models2}), which are low-mass stars that avoid the AGB phase of evolution. This theory has also been explored by \citet{Macdonald_omega_centauri_massloss}, who estimated the integrated RGB and AGB mass losses in the GC $\omega$ Centauri with a combination of empirical calculations and HB models.

The other possibility is that our measured EAGB masses are incorrect. The low masses could be a result of systematic discrepancies in the scaling relations for EAGB stars. Some studies have measured the solar-like oscillations of post-red clump stars that are moving towards the EAGB \citep[e.g.][]{possible_agb_stars,Mosser_stellar_evolstage_classification_seismo}. The stellar luminosity and effective temperature of AGB stars are similar to the RGB, making it difficult to distinguish between these two phases in CMDs. Consequently, EAGB stars are often grouped with RGB stars in asteroseismic studies \citep[e.g.][]{No_AGB1, APOKASC-2}, and can be seen as over-densities in seismic Hertzsprung-Russel diagrams. Recently, \citet{Dreau_catalog} detected differences in the seismic signatures of the helium second-ionisation zones between RGB and EAGB stars, and were therefore able to separate the evolutionary phases of a sample of field stars. However, because these stars are across a wide distribution of masses and metallicities, the scaling relations cannot be robustly tested. The tightness of the CMDs of GCs makes it easier to distinguish between the RGB and EAGB evolutionary phases with photometry. Our study presents the first detections of solar-like oscillations in GC EAGB stars. Further work is needed to study how the thinner convective envelopes of EAGB stars would affect the seismic mass scale.
\subsection{Non-standard evolution for the outlier stars?}
\label{sec:non_standard_evol}
We analysed the reddening dependency on the photometry and mass estimates in Section~\ref{sec:mass_outliers} for the six mass outliers, finding that we require spectroscopic $T_{\text{eff}}$~in order to determine the effect of reddening on the masses. If we assume that the measured masses are correct, then these outlier masses could instead be explained in the context of non-standard evolutionary events.  

Blue stragglers (BS) are massive stars located above the main sequence turn-off and are thought to be the result of either stellar mergers of two main sequence stars \citep{Mergers_BSs} or mass-transfer events in binary systems \citep{MassTransfer_BSs}. First observed in the GC M3 \citep{BSS_first_discovered}, BSs have now been found to be common objects in all GCs, numbering between 40-400 in each cluster \citep{nos_of_BSS1, nos_of_BSS2}. These stars have been modelled to follow the evolution of standard $\sim1.2-1.6~\msun$ stars, and follow an evolutionary track that is slightly bluer than the $\sim0.8~\msun$ RGB in a CMD \citep{eBSS_models1,eBSS_models2}. However, evolved BSs are practically indistinguishable from `normal' GC evolved stars in photometry and will only explicitly stand out in mass. Attempts to identify evolved BSs have used asteroseismology to identify candidates in open clusters \citep{seismo_eBSs_obs1,seismo_eBSs_obs2,2017NGC6819,Brogaard_non_standard_evol_seismo}. Due to the slightly hotter temperatures and larger masses of M4RGB36 and M4RGB217, they could be `hidden' evolved BSs in our RGB sample. M4AGB58 could also be an evolved BSs in the post core He burning phase. 

The estimated mass of a BS star on the main sequence is $\sim1.2-1.6~\msun$. For our over-massive RGB stars with masses of $\sim1~\msun$, this would imply a mass loss of $0.2-0.6~\msun$ before the end of the RGB. The mass of $0.75~\msun$ for our M4AGB58 star would then suggest that a further $~0.25~\msun$ would be lost before the post core He burning phase of BSs. This is a large amount of mass to lose in the short time between the two phases, and suggests that these outliers are not massive enough to be evolved BSs. However, the evolution of BSs is not well understood, and there could be other unknown channels of mass loss. 

The under-massive RHB stars have masses of $\sim 0.53~\msun$. This is consistent with GC white dwarf masses \citep{GC_WD_masses}. Further, an HB star with such a low mass would be found on the extreme end of the blue HB \citep{HB_models2}, not on the RHB. Thus we conclude that there must be inaccuracies in one or more of their stellar or seismic parameters for the under-massive stars, and we cannot speculate further without more data.


\section{Conclusions \& Future Prospects}

With the crowding in the field of view, short-duration observation, and inherent instrumental noise in the telescope, the M4 photometric data is at the limit of what can be successfully analysed from the \textit{K2} mission. By developing our own custom detrending pipeline, we were successful in measuring solar-like oscillations in 75 evolved stars in M4. This is a significant increase compared to the previous sample of 8 stars that were seismically studied by M16. These asteroseismic measurements provide an excellent opportunity to study stellar evolution for low mass/metallicity stars. 

From our mass measurements, we determined a total integrated RGB mass loss of $\Delta\overline{M}=0.17\pm0.01~\msun$, comparing well to the expected $\sim0.2~\msun$ RGB mass loss for GCs from models and empirical determinations \citep[e.g.][]{Macdonald_omega_centauri_massloss,infrared,tucan_47}. By comparing our measured mass loss to stellar models, we estimated that this would correspond to a modelled Reimers' scaling parameter value of $\eta_R = 0.39$, consistent with the \citet{mcdonald_mass_loss_reimers} study. We emphasise the excellent agreement of our initial mass, RHB mass, $\eta_R$, and total integrated RGB mass-loss with previous studies, and also the higher precision of our results by using asteroseismology on such a large sample of stars.   

We searched our mass distributions for  bi-modal mass signatures, which could correspond to the two chemical sub-populations in M4. We did not detect a bi-modal distribution in our RHB sample, in agreement with expectations that the RHB contains only the first sub-population. However, we did observe a tentative bi-modal signature in our RGB sample, which is quantitatively consistent with previous M4 sub-population models. We require light chemical abundances for sub-population membership to confirm this result.

We report the first detections of solar-like oscillations in GC EAGB stars. We found a mean mass for the EAGB phase of $0.54\pm0.01$ that was significantly lower than predicted by standard stellar models ($0.64~\msun$), implying significant mass loss during the HB phase. This contradicts current theories that there is little to no mass loss during the core-helium burning phase for low mass stars. Our average estimate for the EAGB is also consistent with the expected white dwarf mass of $0.5-0.55~\msun$ \citep{GC_WD_masses}. We propose that these stars will likely not reach the thermal pulsing stage and will soon evolve to become white dwarves. Alternatively, there could be unknown systematics in the seismic scaling relations for EAGB stars that could be lowering our mass estimates. Further studies are needed to confirm or deny this.

During the mass analysis, three apparently over-massive stars and three apparently under-massive stars were discovered. We discussed the possibility that the reddening corrections used to estimate the photometric $T_{\text{eff}}$ and bolometric luminosities were inaccurate. We think it is likely that this uncertainty caused mass errors of this magnitude, and we concluded that dust-independent spectroscopic $T_{\text{eff}}$ are necessary to decrease the mass uncertainties. 

Alternatively, assuming the outlier masses are real, we considered scenarios where these stars could be a result of non-standard evolutionary events. We speculated that the over-massive stars could be evolved BSs. For the under-massive stars, we determined that such low mass HB stars would not be found on the RHB, since they should have much higher temperatures, and therefore be on the extreme blue HB. We stress that poor reddening corrections are a more likely explanation for all the outlier stars.

The main uncertainties in this study were in the asteroseismic parameters, $\nu_{\text{max}}$ and $\Delta\nu$. Only a new photometric space mission providing better data will improve this. The parameter with the next largest contribution to the mass uncertainty is $T_{\text{eff}}$. For M4 the uncertainty in $T_{\text{eff}}$ is primarily driven by the inaccurate differential reddening corrections. This can be ameliorated by using spectroscopic $T_{\text{eff}}$, since this removes the dust dependency. We endeavour to acquire spectroscopic temperatures -- and chemical abundances to do sub-population membership -- for our entire sample in the near future at the \textit{Anglo Australian Telescope}.

In this study, we demonstrated the exciting potential of applying asteroseismology to a GC to understand stellar evolution. It is essential that we analyse other GCs using this same method. Unfortunately, the other GCs observed in \textit{K2} cannot be used for RGB mass loss studies, because the HB stars are too faint. Due to this, and also to improve the SNR in the power spectra, we need photometric missions to observe GCs for longer durations. The all-sky surveys of TESS \citep{tess} and PLATO \citep{PLATO} are not optimal for GCs, because of their large pixel sizes ($21$''/pixel and $15$''/pixel respectively), which leads to increased blending between stars in dense fields. A call for an independent mission focused on clusters has been made by the science collaboration `High-precision AsteroseismologY of DeNse stellar fields' \citep[HAYDN;][]{HAYDN}. They are proposing a dedicated space mission committed to gathering long-period and high-precision data of stellar clusters, to study stellar evolution further with asteroseismology.



\section*{Acknowledgements}

We thank Yazan Momany for providing M4 UBVI photometry, and Benjamin Hendricks for providing the M4 reddening map.

S.W.C. acknowledges federal funding from the Australian Research Council through a Future Fellowship (FT160100046) and Discovery Project (DP190102431). D.S. is supported by the Australian Research Council (DP190100666). Parts of this research was supported by the Australian Research Council Centre of Excellence for All Sky Astrophysics in 3 Dimensions (ASTRO 3D), through project number CE170100013. This research was supported by use of the Nectar Research Cloud, a collaborative Australian research platform supported by the National Collaborative Research Infrastructure Strategy (NCRIS).

This publication makes use of data products from the Two Micron All Sky Survey, which is a joint project of the University of Massachusetts and the Infrared Processing and Analysis Center/California Institute of Technology, funded by the National Aeronautics and Space Administration and the National Science Foundation. This paper includes data collected by the Kepler mission. Funding for the Kepler mission is provided by the NASA Science Mission directorate. This work has made use of data from the European Space Agency (ESA) mission Gaia (https://www.cosmos.esa.int/gaia), processed by the Gaia Data Processing and Analysis Consortium (DPAC, https://www.cosmos.esa.int/web/gaia/dpac/consortium). Funding for DPAC has been provided by national institutions, in particular the institutions participating in the Gaia Multilateral Agreement.

This research made use of Lightkurve, a Python package for Kepler and TESS data analysis \citep{Lightkurve}. This work was also made possible by the following open-source \texttt{PYTHON} software: \texttt{asfgrid} \citep{Sharma_asfgrid}, \texttt{pySYD} \citep{pySYD_chontos}, \texttt{NumPy} \citep{numpy}, \texttt{pandas} \citep{pandas}, and \texttt{Matplotlib} \citep{matplotlib}.

We thank the referee, L\'eo Girardi, for a careful reading and constructive comments on the manuscript.

We wish to acknowledge the people of the Kulin Nations, on whose land Monash University operates, and in which the majority of this research was conducted. We pay our respects to their Elders, past, and present. 
\section*{Data Availability}
The main data in this article (Table~\ref{tab:final_results}) will be available in the online only supplementary material and an online catalog. The light curves will be shared on reasonable request to the author. 
 



\bibliographystyle{mnras}
\bibliography{refs}








\bsp	
\label{lastpage}
\end{document}